\documentclass[lettersize,journal]{IEEEtran}
\usepackage{amsmath,amsfonts}
\usepackage{algorithmic}
\usepackage{array}
\usepackage[caption=false,font=normalsize,labelfont=sf,textfont=sf]{subfig}
\usepackage{textcomp}
\usepackage{stfloats}
\usepackage{url}
\usepackage{verbatim}
\usepackage{graphicx}
\def\BibTeX{{\rm B\kern-.05em{\sc i\kern-.025em b}\kern-.08em
    T\kern-.1667em\lower.7ex\hbox{E}\kern-.125emX}}
\usepackage{balance}

\usepackage{bm}
\usepackage{pifont}
\usepackage{booktabs}
\usepackage{amssymb}
\usepackage{footnote}
\usepackage{multirow}
\usepackage{enumerate}

\newcommand{\myref}[1]{Equation \ref{#1}}
\newcommand{\myfigure}[1]{Fig. \ref{#1}}
\newcommand{\myTable}[1]{Table \ref{#1}}
\newcommand{\myPara}[1]{Section \ref{#1}}

\begin{document}

\title{Rumor Detection with Hierarchical Representation on Bipartite Adhoc Event Trees}

\author{Qi~Zhang,
        ~Yayi~Yang,
        ~Chongyang~Shi,
        ~An Lao,
        ~Liang~Hu,
        ~Shoujin~Wang,
        and~Usman~Naseem
\IEEEcompsocitemizethanks{
\IEEEcompsocthanksitem Qi Zhang is with Data Science Lab, University of Technology Sydney, Ultimo, NSW 2007, Australia and Beijing Institute of Technology, Beijing, 100081, China. E-mail: qi.zhang-13@student.uts.edu.au
\IEEEcompsocthanksitem Yayi Yang, Chongyang Shi (corresponding author), and An Lao are with the School of Computer Science, Beijing Institute of Technology, Beijing, 100081, China. E-mail: \{yayi.yang, cy\_shi, an.lao\}@bit.edu.cn
\IEEEcompsocthanksitem Liang Hu is with Tongji University, Shanghai, 200092, China and DeepBlue Academy of Sciences, Shanghai, 200336, China. E-mail: rainmilk@gmail.com
\IEEEcompsocthanksitem Shoujin Wang is with Data Science Institute, University of Technology Sydney, Sydney,  NSW 2007, Australia. E-mail: shoujin.wang@uts.edu.au
\IEEEcompsocthanksitem Usman Naseem is with University of Sydney, Sydney, Camperdown, NSW 2006, Australia. E-mail: usman.naseem@sydney.edu.au
}
}

\markboth{Journal of \LaTeX\ Class Files,~Vol.~18, No.~9, September~2020}%
{How to Use the IEEEtran \LaTeX \ Templates}

\maketitle
\begin{abstract}
The rapid growth of social media has caused tremendous effects on information propagation, raising extreme challenges in detecting rumors. Existing rumor detection methods typically exploit the reposting propagation of a rumor candidate for detection by regarding all reposts to a rumor candidate as a temporal sequence and learning semantics representations of the repost sequence. However, extracting informative support from the topological structure of propagation and the influence of reposting authors for debunking rumors is crucial, which generally has not been well addressed by existing methods. In this paper, we organize a claim post in circulation as an adhoc event tree, extract event elements, and convert it to bipartite adhoc event trees in terms of both posts and authors, i.e., author tree and post tree. Accordingly, we propose a novel rumor detection model with hierarchical representation on the bipartite adhoc event trees called BAET. Specifically, we introduce word embedding and feature encoder for the author and post tree, respectively, and design a root-aware attention module to perform node representation. Then we adopt the tree-like RNN model to capture the structural correlations and propose a tree-aware attention module to learn tree representation for the author tree and post tree, respectively. Extensive experimental results on two public Twitter datasets demonstrate the effectiveness of BAET in exploring and exploiting the rumor propagation structure and the superior detection performance of BAET over state-of-the-art baseline methods.
\end{abstract}


\section{Introduction}\label{Introduction}
\IEEEPARstart{W}{ith} the rapid development of social platforms, social media has become a convenient and essential tool for people to release and access news, comments, and speak out. However, social media can be a double-edged sword, as it facilitates the release and rampant spread of large amounts of fake or unverified information, leading to tremendous adverse effects on our society~\cite{2016The, Zhiying2020Bibliometric}. For example, widespread rumor-mongering has seriously impacted political decision-making and manipulated public opinions in the 2016 US presidential election~\cite{2018The}. During the COVID-19 public health crisis, the wide-ranging rumor that oral disinfectants can prevent infection and that heating masks can kill the virus led to the emergence of a large-scale health crisis~\cite{abs-2104-12556}. Thus, detecting rumors is urgent and beneficial to avoid their potential threat and negative influences on social media and society.
 
\begin{figure}[!t]
 	\centering
 	\includegraphics[width=0.75\linewidth]{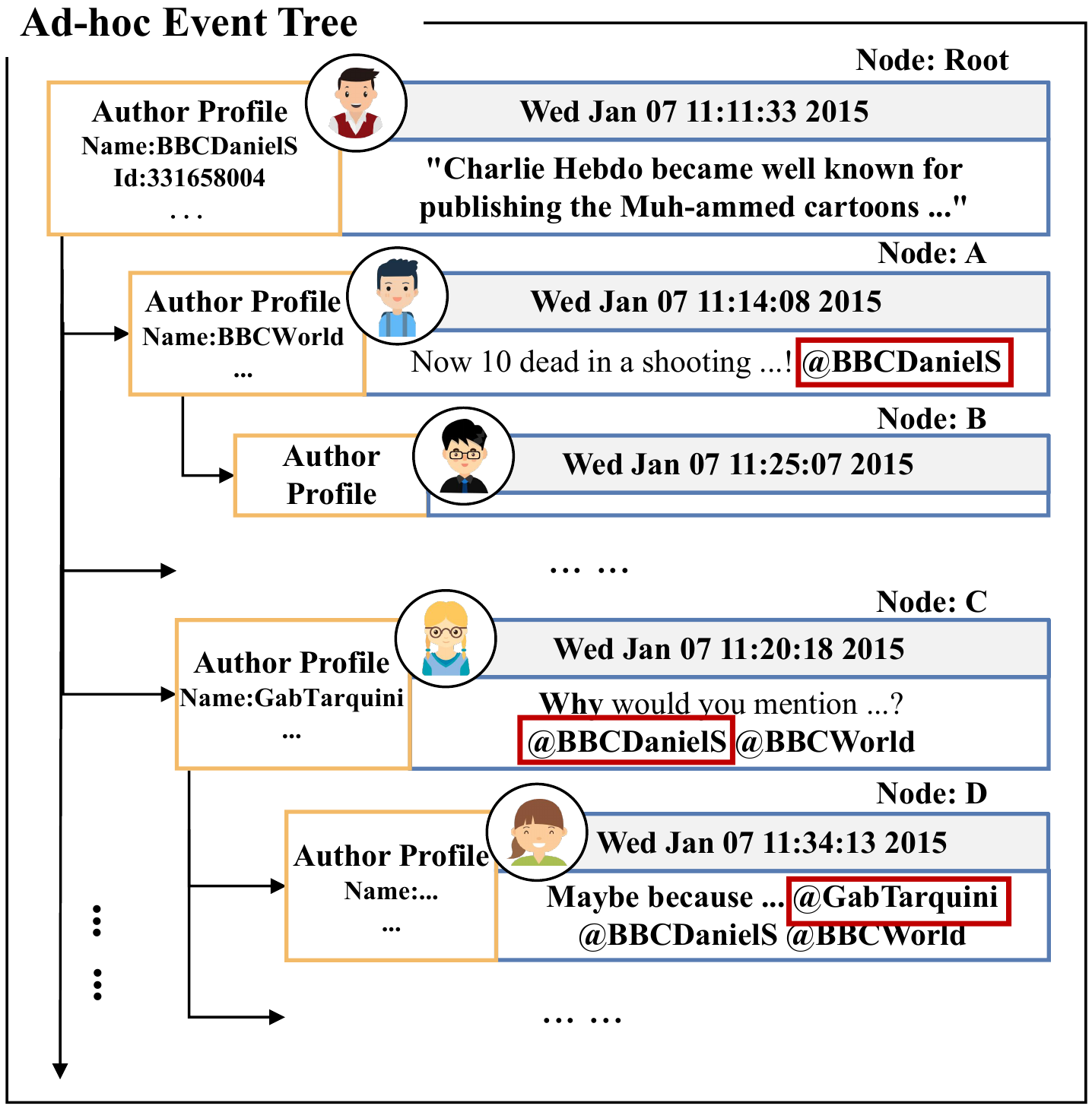}
 	\caption{The tree shows an adhoc event in circulation which is derived from the claim post \textit{Charlie Hebdo headquarters being attacked by armed men}. In the tree, each node denotes a tuple related to the event, including a post related to the event, the author profile, timestamp of the post. Each arrow denotes one repost operation. The root node denotes the claim post and all the others denote reposted posts. The ``@" in the red box forms the repost.
 	}
 	\label{fig:problem}
\end{figure} 

The sociological definition of a rumor is a circulating piece of unverified information (e.g., post)~\cite{ZubiagaABLP18,Vega-OliverosZR22}. The task of rumor detection aims to verify the information veracity and distinguish rumors from non-rumors~\cite{9594513}. Dominant methods treat rumor detection as a binary classification task and solve the task referring to the three key properties of rumors, i.e., post content, author, and temporal propagation~\cite{ZubiagaABLP18}. For example, traditional methods adopt handcrafted
features of text contents, user characteristics, and propagation patterns to train a classifier based on decision tree~\cite{CastilloMP11} or support vector machine~\cite{KwonCJCW13}. Deep learning methods extract semantic features of stance or attitude from post contents~\cite{LukasikBCZLP19,9594513,TanCKZAS22} or capture high-level representations of sequential features and structural information from propagation paths~\cite{BianXXZHRH20,LaoSY21} to make the classification. 

Both traditional and deep learning methods conclude that the temporal propagation of a rumor, in addition to its inherent information (i.e., its post and author), is effective for verifying the veracity of the rumor. Besides, increasingly prevailing methods pay more attention to the auxiliary information along with propagation and learn informative representation from the rumor propagation structure for rumor detection. These methods can be roughly divided into two categories: (1) methods modeling rumor propagation from the chronological order perspective and (2) methods modeling rumor propagation from a topological perspective. For example, RNN-based models~\cite{DBLP:conf/ijcai/MaGMKJWC16, DBLP:conf/pakdd/ChenLYZ18} arrange claim posts and comments in chronological order and exploit the sequential texts using recurrent neural units and attention mechanism for rumor detection. These models simplify posts' propagation and ignore the topological structure in the propagation. To bridge this gap, some traditional models, e.g., graph-kernel-based methods~\cite{DBLP:conf/icde/WuYZ15, DBLP:conf/acl/MaGW17}, and recent neural models, e.g., tree-based RvNN~\cite{DBLP:conf/acl/WongGM18} and graph-based Bi-GCN~\cite{DBLP:conf/aaai/BianXXZHRH20, DBLP:conf/acl/WeiHZYH20}, exploit the topological structure of rumor propagation to learn informative representations for rumor detection. Notably, these methods benefit from the chronological order and topological structure of rumor propagation, thus obtaining more effective information to debunk rumors. This indicates that the structure information of rumor propagation implies important spreading behaviors of rumors.


However, the above methods generally model the structure of the rumor propagation path from the perspective of posts while ignoring authors during rumor propagation. Intuitively, the veracity of a rumor depends on not only its post content but also its propagation information, including the semantics (e.g., stance and attitude) of posts and the influence of (reposting) authors (e.g., reputation and reliability). Quite a few existing methods leverage authors' influence to enhance rumor detection. These methods~\cite{DBLP:conf/acl/WuRZLN20,DBLP:conf/acl/LuL20} typically treat author information as additional information to the post content to improve the presentation of the text content of the post. Considering the authors' influence, we treat a claim post of a rumor or non-rumor in circulation as an adhoc event, including the authors (who), the time sequence (when), and the posts (what) in circulation. Each adhoc event is organized with a tree to describe the propagation path of the information, which is denoted as an adhoc event tree as shown in the left on \myfigure{fig:problem}. We argue that author information, e.g., profiles, which implies his/her reliability, also play an indispensable role in judging the veracity of his/her claims. For example, a verified author is indicated by the attribute "verified", and an author with a large number of followers may have a high probability of posting a true claim. To this end, it is necessary and reasonable to exploit author information in the adhoc event to judge the veracity of the corresponding posts. In addition, as the saying goes that "birds of a feather flock together", it is intuitively effective to exploit the information of authors in circulation to verify the veracity of their posts and support the reliability of the author of the claim post.

\begin{figure}[!t]
	\centering
	\includegraphics[width=\linewidth]{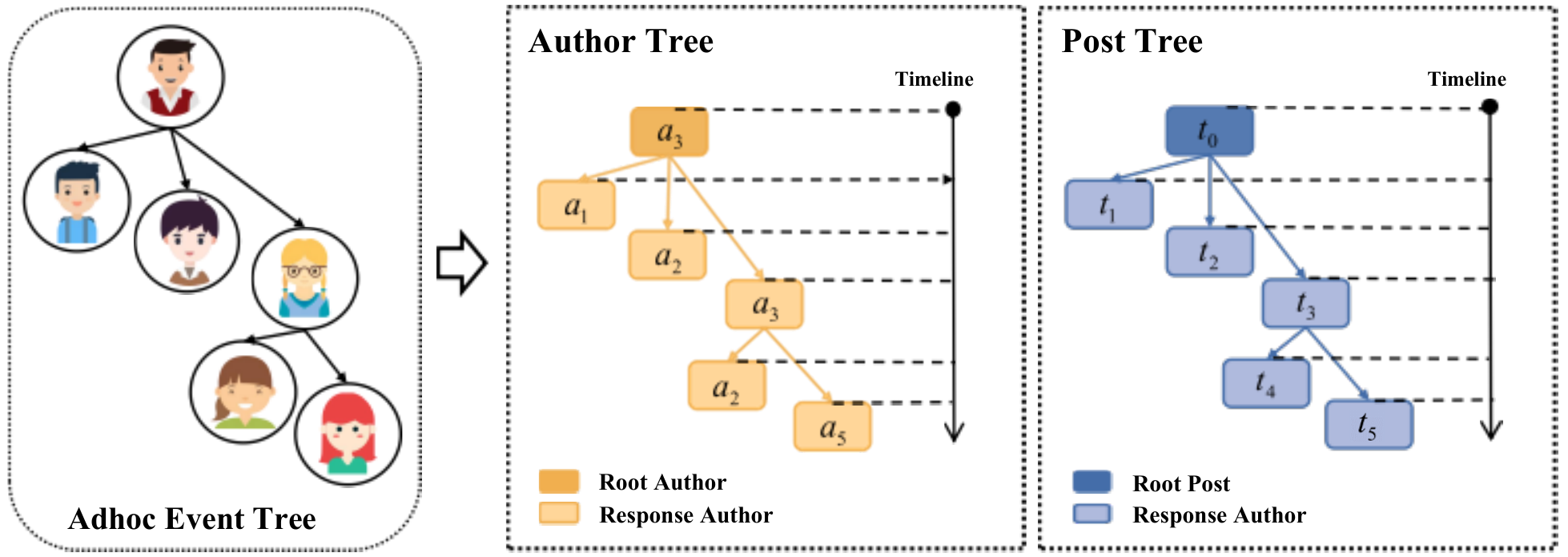}
	\caption{The propagation structure of the adhoc event tree (left) in \myfigure{fig:problem} and the corresponding converted bipartite adhoc event trees (right). The author (post) tree has the same structure as the original adhoc event tree. In the author (post) tree, each node denotes an author (post), and all nodes are arranged in chronological order from top to bottom.}
	\label{fig:bipartite}
\end{figure}


In light of the above discussion, we convert the adhoc event tree as shown in \myfigure{fig:problem} into bipartite adhoc event trees in terms of both posts and authors, denoted as \textit{author tree} and \textit{post tree} respectively, as shown in \myfigure{fig:bipartite}. Based on the obtained bipartite trees, the major challenge for debunking rumors is to obtain sufficient information on both posts and authors from the post tree and author tree, respectively. This triggers two challenges (a) how to learn an informative representation for each of the tree nodes, and (b) how to effectively capture the structural correlations between the root post (author) and the response posts (authors) within the bipartite adhoc trees to represent the propagation structure? Accordingly, we propose a novel rumor detection model which learns hierarchical representation from the bipartite adhoc event trees called BAET. The model introduces word embedding and feature encoder for author and post tree, respectively, to learn the embedding of posts and authors, and a root-aware attention module to learn node representation for both posts and authors to address the challenge (a). Motivated by \cite{DBLP:conf/acl/WongGM18}, we further introduce the tree-like RNN model (i.e., RvNN) to capture the aforementioned structural correlations and propose a tree-aware attention module to learn tree representation for the author tree and post tree respectively to address the challenge (b). Our contributions can be summarized as follows:
\begin{itemize}	
	\item We treat a claim post in circulation as an adhoc event from which we construct bipartite (post/author) adhoc trees to provide additional author information for rumor detection. To achieve this, we propose a novel rumor detection model called BAET, utilizing signals from both posts and authors to debunk rumors effectively.
	\item We introduce word embedding and feature encoder to embed post and author, respectively, and design a root-aware attention module for learning node-level representation effectively.
	\item We adopt RvNN to learn the structural correlations over nodes within trees and then propose a tree-aware attention module to adaptively learn the structural-level representation from the bipartite adhoc trees.
	\item Extensive experimental results show that BAET achieves significant improvements compared with the state-of-the-art rumor detection models on public rumor detection datasets PHEME and RumorEval, which proves the contribution of author information to rumor detection.
\end{itemize}


\section{Related Work} \label{Related Work}
 
\subsection{Early rumor detection based on feature engineering}
Researchers in different disciplines have investigated rumor detection, including philosophy and humanities~\cite{2007Rumor}, social psychology~\cite{1980Who,2005Rumor}, computer science and artificial intelligence~\cite{DBLP:conf/emnlp/QazvinianRRM11,DBLP:conf/icwsm/HannakMKW14}. It is difficult to identify rumors early on because a large number of news investigations are required to verify suspicious tweets. With the increased usage of social media, there is a massive information overload, making rumor detection a time-consuming and labor-intensive task. Early studies focused on using manual linguistic features extracted from rumor textual contents (source/root and responsive posts) to decide rumor labels (i.e., true or false) via traditional machine learning methods~\cite{DBLP:conf/www/ZhaoRM15, 2015Automatic, DBLP:conf/semeval/ChenLK17}. For example, Zhao et al. \cite{DBLP:conf/www/ZhaoRM15} believed that people exposed to a rumor claim would seek more information or express skepticism before deciding whether to believe it or not. Thus, he clustered the claim and its successor texts which contain doubtful expressions such as ``Is this true?" and ``Really?", and then used classical classification models such as SVM and decision tree to conduct the classification task. Ma et al.~\cite{DBLP:conf/acl/MaGW17} aggregated all of the responses to the root claim to form a rumor propagation tree and then identified the claim using SVM with different kernel numbers. 
 
These methods rely on manually created features, which can not guarantee the accuracy of rumor classification. To achieve real-time tracking and identification of rumors, researchers must shift their focus to automatic rumor detection methods. The initial research on the automatic rumor detection method is based on text mining with supervised models built on feature engineering. For instance, Ma et al.~\cite{DBLP:conf/cikm/MaGWLW15} proposed a dynamic time series structure (DSTS) that can capture a variety of social context text features based on time series changes, as well as the various features generated by rumors along with time, allowing rumors to be detected in different periods. Rath et al.~\cite{DBLP:journals/snam/RathGMS18} proposed a new trustworthiness concept based on responsive post characteristics to identify authors who spread rumors automatically. Because of the large amount of noise information, such feature engineering-based methods extract more single or coarse-grained features, and the accuracy of the classification task cannot be guaranteed.
 
 \subsection{Automatic rumor detection based on deep learning}
 Deep learning research solves the automatic feature extraction problem while achieving fine granularity and diversity. Researchers are no longer limited to the single processing and use of features because of richer and more accurate feature information. However, they are instead exploring the fusion of multiple features and structural modeling.
 
\textbf{Serialized propagation topology representation.} 
Some methods place less emphasis on the propagation topology and instead focus on propagation time sequence properties. These researchers hoped to explore relevant clues generated over time to characterize the different characteristics of the rumor at different times by modeling the propagation process of information with time characteristics. Yu et al.~\cite{DBLP:journals/compsec/YuLWWT19} designed a CNN-based model called CAMI to extract flexible content sequence between key characteristics, formation of high-level interactions, and aids in the effective classification. Extensive mixed features, such as time series and content-related features, are increasingly important to research. Thus Ma et al.~\cite{DBLP:conf/ijcai/MaGMKJWC16} used an RNN network to process each time step in the rumor propagation sequentially. Chen et al.~\cite{DBLP:conf/pakdd/ChenLYZ18} proposed an attention mechanism to capture longer-term dependencies and enhance the long-term temporal features~\cite{ZhangCSN22}. In addition, attention networks~\cite{DBLP:conf/acl/WeiHZYH20} and Transformers~\cite{DBLP:conf/aaai/KhooCQ020} were adopted to better mine the posts' features with time sequences as a clue to identify rumors.
 
Author traits, on the other hand, are frequently used to define authors' personality information, such as authors' profile information and social relationships, which are used to vividly describe the authors' involvement in propagation and their behavior~\cite{DBLP:conf/icde/WuYZ15}. Shu et al.~\cite{DBLP:conf/asunam/ShuZWZL19} argued that fake news could imitate real news, which means that the dissemination of authors' information must be thoroughly researched. Ruchansky et al.~\cite{DBLP:conf/cikm/RuchanskySL17} developed a tri-relationship embedding framework TriFN that can learn the characteristics of forwarded texts and author profiles while also generating credibility scores to help with rumor detection. Shu et al.~\cite{DBLP:conf/wsdm/ShuWL19} built five novel metrics to interactively model the relationships between news and users (publishers, social participants, etc.). Recently, Wu et al.~\cite{DBLP:conf/acl/WuRZLN20} constructed a decision tree mechanism to address rumor interpretability issues, using users' credibility scores as evidence to help identify rumors. Lu et al.~\cite{DBLP:conf/acl/LuL20} built a graph-aware joint attention network to detect fake news by taking into account the root article's text content (claim), forwarding author sequence (sequential propagation) and author information (author profile). These methods, however, cannot accurately reflect the characteristics of actual rumor propagation and spreading because they focus on mining the characteristics of authors and posts changing over time via simple temporality while ignoring the internal topology of rumor propagation.
 
\textbf{Structured propagation topology representation.} The rumor propagation topology representation is concerned with modeling the rumor propagation structure, gathering various information such as stances and sentiments in the topology, and mining the interactive information of multiple features of posts and authors to enrich the rumor representation. Simultaneously, understanding the process and mode of rumor propagation aids in identifying rumors at an early stage of rumor propagation and achieves a good explanatory effect in detecting the authenticity of rumors. Rumor propagation topology enables us to better abstract rumor propagation patterns and further excavate the correlations between hybrid features via propagation aggregation~\cite{LiuXWC15,DBLP:journals/csur/ZhouZ20,DBLP:conf/bigdataconf/WangT15}, resulting in good interpretability on the veracity of rumors. Wu et al.~\cite{DBLP:conf/icde/WuYZ15} calculated the substructural similarity of two propagation trees on Sina Weibo datasets for detection. Similarly, Ma et al.~\cite{DBLP:conf/acl/MaGW17} employed the tree kernel to compute similar substructures and identify various types of rumors in Twitter datasets. The above two methods rely on the similarity of the two trees. But they cannot distinguish between individual trees and are still based on feature engineering rather than deep learning. Moreover, considering that RvNN \cite{DBLP:conf/icml/SocherLNM11, article} were originally used to learn phrase or sentence representations for semantic parsing, Wong et al.~\cite{DBLP:conf/acl/WongGM18} proposed two types of recursive tree-structured neural network models to learn representations from their structure and contents. Ma et al.~\cite{DBLP:journals/tist/MaGJW20} recently developed an improved rumor detection model based on Wong et al.~\cite{DBLP:conf/acl/WongGM18} with designed discriminative attention mechanisms for model optimization.
 
Bian et al. \cite{DBLP:conf/aaai/BianXXZHRH20} believed that Wong et al.~\cite{DBLP:conf/acl/WongGM18} only considered propagation depth, ignoring spread. Two graph convolutional networks (GCN) were used to learn the top-down propagation depth and the down-top disperse breadth simultaneously. Wei et al.~\cite{DBLP:conf/acl/WeiHZYH20} is also based on GCN networks and considers the reliability of the rumor propagation relationship when optimizing the propagation model for a better rumor classification effect. Graph attention networks (GAT) are also used in various research fields. Lin et al.~\cite{DBLP:conf/emnlp/LinMCYCC21} proposed a new ClaHi-GAT model to represent the action graph between the post content and the author and learn the characteristics of rumors by the multi-layer attention mechanism of the root post layer and event layer. In addition, Liu et al.~\cite{9750388} proposed a novel rumor detection framework based on structure-aware retweeting GNNs based on a converted tractable tree. Wei et al.~\cite{9837882} further considered the uncertainty of interactions caused by users’ various subjective factors and proposed fuzzy graph convolutional networks to model uncertain interactions in the information cascade of a tweet graph. Although such methods are constantly improving the modeling of rumor propagation topological structure to obtain better rumor representations, most ignore author-level information, including simple user personality characteristics (author profiles) and the structured interaction relationship formed between users during the spread process.
 
\section{Problem Statement} \label{Problem Statement}
In this section, we introduce the problem statement of rumor detection in which the task is to learn a classifier to distinguish a claim post of a rumor or non-rumor. To achieve this, we treat the claim post with its responsive posts in circulation as an adhoc event tree and build bipartite adhoc event trees consisting of a post tree and an author tree correspondingly to extract more informative supports for detection. Formally, let $\Gamma=\{{t_0,t_1,t_2,...,t_n}\}$ be a sequence of posts in a post tree in chronological order where $n$ denotes the number of responsive posts. Herein, $t_0$ corresponding to a claim post is the root node of the post tree, $t_k (k>0)$ in $\Gamma$ corresponds to the $k$-th responsive posts, and $\{{t_1,t_2,...,t_n}\}$ denotes the sequence of re-posts. Correspondingly, we denote the sequence of authors in an author tree as $\Theta=\{{a_0,a_1,a_2,...,a_n}\}$, where each author $a_i$ published the claim $t_i$ and is involved with various features, such as the number of followers and friends, as shown in \myTable{userFeature}. We further build the responsive connections between posts (authors) within the post (author) sequence from the "forward" and "reply" operations revealed by the sign ``@" in the posts, depicting the propagation structure of the post (author) tree. Accordingly, a classifier $C$ is then build to predict the label $y$ of the claim $t_0$ associated with the inputs of $\Gamma$ and $\Theta$ where $y\in\{0,1\}$ and 0 indicates a rumor otherwise 1, that is $C: (t_0,\Gamma,\Theta)\to y$. Allowing for obtaining propagation information, we revisit the critical edge-directional propagation problem in the tree and treat the direction of each arrowed connect between posts (authors) indicating the direction of information transmission. 
 
\section{The BAET Model} \label{Method}
\begin{figure*}[htbp]
	\centering
	\includegraphics[width=0.95\linewidth]{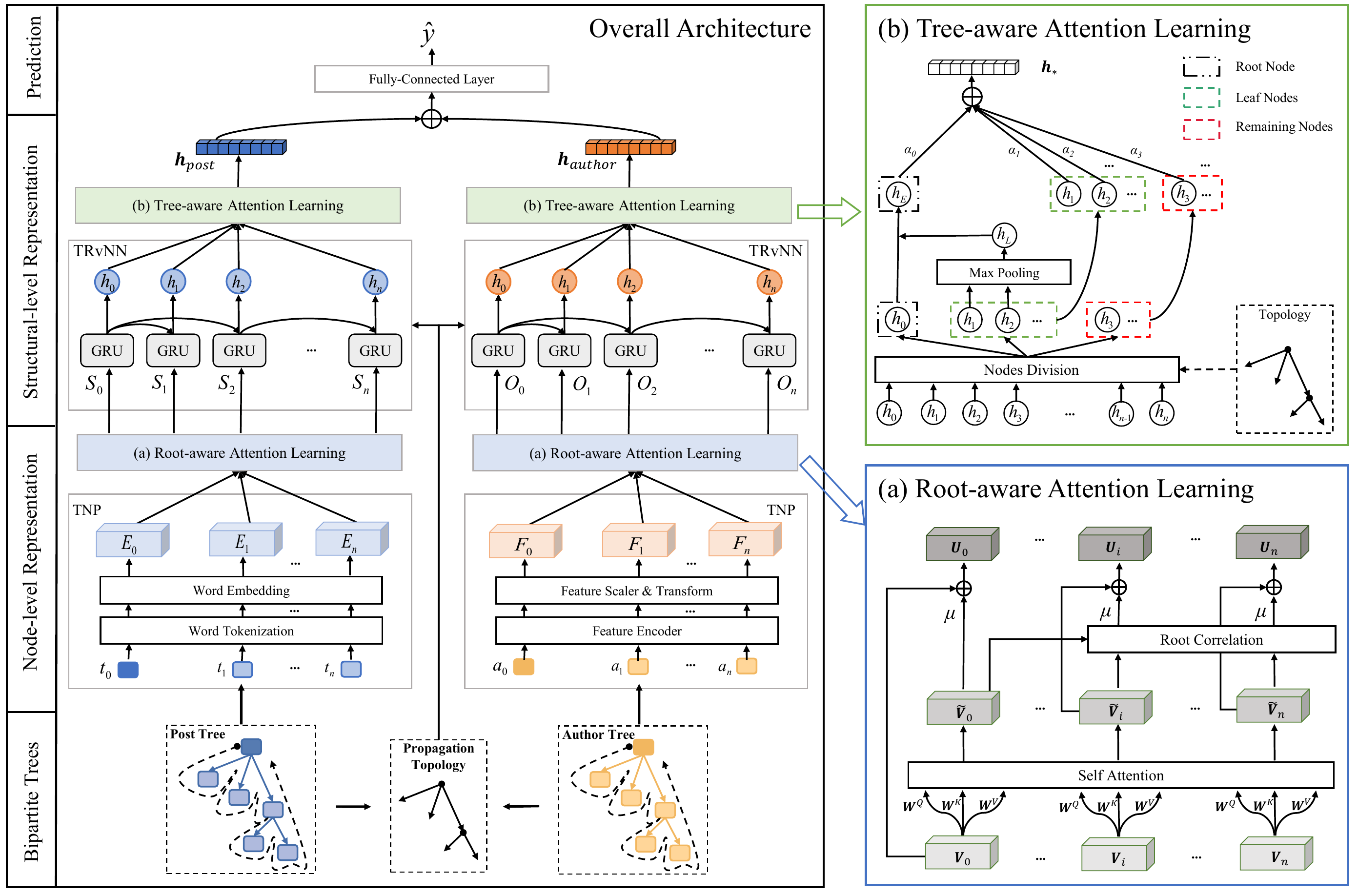}
	\caption{The overall architecture of BAET. It comprises three key modules: \emph{Node-level Representation}, \emph{Structural-level Representation}, and \emph{Prediction} modules. 1) Node-level representation module includes two sub-modules of \textit{Tree Node Process} (TNP) and \textit{Root-aware Attention Learning} (TAL). 2) Structural-level representation module contains two sub-modules of \textit{Tree Variant RNN} (TRvNN) and \textit{Tree-aware Attention Learning} (TAL). 3) Prediction layers are adopted to classify rumors and non-rumors. Both node-level representation modules and structure-level representation modules are applied to the post tree and author tree to learn more informative representations of post nodes and author nodes, respectively.}
	\label{fig:myMod}
 \end{figure*}

We introduce our rumor detection model BAET in this section. The overall architecture of BAET is depicted in \myfigure{fig:myMod}, containing three key components: \emph{Node-level Representation Module}, \emph{Structural-level Representation Module}, and \emph{Prediction Module}. \myTable{notation} summarizes the key notations that are used throughout this paper.


\begin{table}[!t]
  \centering
  \caption{Primary Notation Descriptions}
    \begin{tabular}{l|l}
    \toprule
    \midrule
    Notation & Explanation \\
    \midrule
    $t_0$    & The claim post text \\
    $t_i$    & The $i$-th responsive post text \\
    $a_0$    & The claim author \\
    $a_i$    & The $i$-th responsive author \\
    $\mathbf{E}_i$    & The embedding matrix of $i$-th post text \\
    $\mathbf{F}_i$    & The feature representation matrix of $i$-th author \\
    $\mathbf{h}_i$    & The $i$-th hidden representations of GRU Cell \\
    \midrule
    \bottomrule
    \end{tabular}%
  \label{notation}%
\end{table}%

\subsection{Node-level Representation Module} \label{sec:NRM}
We employ the node-level representation module in both the post tree and author tree to learn a more informative representation of each node from the perspectives of semantics (content) and interactive (propagation) correlations in the bipartite tree. Specifically, this module consists of two sub-modules: \emph{Tree Nodes Process} (TNP) and \emph{Root-aware Attention Learning} (RAL). TNP is introduced to perform embeddings of post nodes and author nodes. The embeddings are then fed into the RAL model to learn the correlations between words/author features within each post/author and the semantic/interactive correlations between each responsive post/author and the claim (root) post/author.

\textbf{(1) Tree Node Process (TNP).} We first introduce \textit{Tree Node Process} in the post tree to capture the text information of all posts in an adhoc event. The topological post tree obtained in \myfigure{fig:bipartite} is processed along the tree structure, where all post text information and topological structure in the tree are recorded. Notably, the topological structure in the tree mainly consists of root nodes, non-leaf nodes, and leaf nodes. We denote the maximum length of the post text as $L_m$ and collect all words in all post texts as a corpus. All words are then tokenized such that each word can be represented by a corpus index. Formally, all the tokenized texts are initialized as  $\{\widetilde{\mathbf{t}}_0,\widetilde{\mathbf{t}}_1,\widetilde{\mathbf{t}}_2,...,\widetilde{\mathbf{t}}_n\}$ with $\widetilde{\mathbf{t}}_i \in\mathbb{R}^{L_m \times 1}$ and mapped into a low-dimensional space through an embedding layer as follows:
\begin{equation}
	\widetilde{\mathbf{E}}_{i}=\operatorname{Embedding}(\widetilde{\mathbf{t}}_i)
\end{equation}
where $\widetilde{\mathbf{E}}_i\in\mathbb{R}^{L_m \times d}, i \in [0, n]$, and $d$ is the dimension of embedding. We further count the frequency of each word appearing in its post and obtain the frequency vectors of all posts as $\{\mathbf{q}_0, \mathbf{q}_1, \mathbf{q}_2,..., \mathbf{q}_n\}$ with the word frequencies of $i$-th post $\mathbf{q}_i \in \mathbb{R}^{1 \times L_m}$ and $i \in [0, n]$. Then, we get the frequency-enhanced embeddings of posts as below:
\begin{equation}
	\mathbf{E}_i=\mathbf{q}_i*\widetilde{\mathbf{E}}_i
\end{equation}
where $*$ denotes the element-wise product, and $\mathbf{E}_i\in\mathbb{R}^{L_m\times d}$. Accordingly, we obtain the embeddings of all the posts in the post tree: $\{\mathbf{E}_0, \mathbf{E}_1,...,\mathbf{E}_n\}$.

\begin{table}[!t]
  \centering
  \caption{Details of Author Features}
    \begin{tabular}{c|l|c}
    \toprule
    \midrule
    \textbf{Category} & \multicolumn{1}{c|}{\textbf{Description}} & \multicolumn{1}{c}{\textbf{Value Range}} \\
    \midrule
    \multicolumn{1}{c|}{\multirow{10}[2]{*}{\begin{tabular}[c]{@{}c@{}}Author basic\\ features\end{tabular}}} & number of followers & int [0, $\infty$) \\
          & number of friends & int [0, $\infty$) \\
          & number of favorites & int [0, $\infty$) \\
          & number of reposts & int [0, $\infty$) \\
          & number of statuses & int [0, $\infty$) \\
          & verified & bool (0/1) \\
          & geo-enabled & bool (0/1) \\
          & time-zone enabled & bool (0/1) \\
          & time of posting & timestamp \\
          & time of creating an account & timestamp \\
    \midrule
    \multicolumn{1}{c|}{\multirow{6}[2]{*}{\begin{tabular}[c]{@{}c@{}}Author writing-habit\\ features\end{tabular}}} & similarity to claim & float [0, 1) \\
          & post text length & int [0, 30] \\
          & max text length (a tree) & int [0, 30] \\
          & average word length & int [0, 20] \\
          & question mark count & int [0, 10] \\
          & exclamation mark count & int [0, 10] \\
    \midrule
    \bottomrule
    \end{tabular}%
  \label{userFeature}%
\end{table}%

Regarding the author nodes, we have two types of author features, i.e., the author's basic features and the authors' writing habits, as shown in \myTable{userFeature}. Specifically, ``similarity to claim" is calculated using Jaccard similarity, expressing the relevance between the repost and the claim post, ``post text length" records the number of words in the post, ``max post text length" is the maximum of text length in the tree, and ``average word length" reveals the author's habits in terms of the average length of words in his/her posts.

We perform feature encoding for each type of feature separately and then adopt one-hot encoding to transform categorical features such as ``verified", ``geo-enabled", and ``time-zone enabled". In addition, the timestamp features (i.e., ``time of posting" and ``time of creating an account" are replaced with timestamp intervals calculated as follows:
\begin{equation}
	\Delta s_i=\log{|s_i - s_0 + 1|}, \ i \in [0, n]
\end{equation}
where $s_0$ and $s_i$ correspond to the timestamps of the root post and the $i$-th responsive post, respectively. Accordingly, we obtain two matrices of raw basic features and writing-habit features for all authors. The two matrices are then normalized to $[0,1]$ using the min-max normalization separately, resulting in two normalized feature matrices $\mathbf{F}^b=[\mathbf{f}^b_0, \mathbf{f}^b_1,...,\mathbf{f}^b_n]$ and $\mathbf{F}^h=[\mathbf{f}^h_0, \mathbf{f}^h_1,...,\mathbf{f}^h_n]$ where $\mathbf{f}^b_i$ and $\mathbf{f}^h_i$ denote the normalized feature vectors of the basic features and writing-habit features of the $i$-th author (i.e., $a_i$), respectively. Finally, we can obtain the feature representation for each author calculated as follows:
\begin{equation}
	\bm{F}_i=\mathbf{f}^b_i\bm{W}^b\oplus \mathbf{f}^h_i\bm{W}^h, \ i \in [0, n]
\end{equation}
where $\bm{F}_i\in\mathbb{R}^{(|\mathbf{f}^b|+|\mathbf{f}^h|)\times d}$ denotes the representation embedding of author $a_i$, $\bm{W}^b \in \mathbb{R}^{1 \times d}$ and $\bm{W}^h \in \mathbb{R}^{1 \times d}$ denote the transform matrices, and $\oplus$ denotes the concatenation operation. Finally, we obtain the feature representations of all authors $\{\bm{F}_0, \bm{F}_1,...,\bm{F}_n\}$.

\textbf{(2) Root-aware Attention Learning (RAL).} We propose a root-aware attention learning module consisting of two attention modules to learn comprehensive node representations for both the post tree and the author tree. As shown in \myfigure{fig:myMod} (a), we use $\{\mathbf{V}_0,\mathbf{V}_1,...,\mathbf{V}_n\}$ to denote the representation matrices obtained from either the post tree or author tree via the tree nodes process module, i.e., $\mathbf{V}_i=\mathbf{E}_i$ or $\mathbf{V}_i=\mathbf{F}_i$. 

First, to learn the correlations between words (features) within a post (author) node, we leverage the self-attention mechanisms to enhance the representations obtained through the \emph{Tree Nodes Process} module to capture the semantics or features correlations within each post/author node:
\begin{equation}
	\widetilde{\mathbf{V}}_i = \mathcal{A}(\mathbf{V}_i\bm{W}^Q,\mathbf{V}_i\bm{W}^K,\mathbf{V}_i\bm{W}^V)
	\label{con:inatt}
\end{equation}
where $\bm{W}^Q,\bm{W}^K,\bm{W}^V\in \mathbb{R}^{d \times d}$ are projection weight parameters, $\mathcal{A}(\mathbf{Q}, \mathbf{K}, \mathbf{V})=\operatorname{Softmax}(\frac{\mathbf{Q}\mathbf{K}^T}{\sqrt{d_k}})\mathbf{V}$ denotes the calculation of self-attention, and $\widetilde{V}_i$ denotes the output of self-attention.

Secondly, with the increase in the number of participants in actual rumor/non-rumor propagation, the claim post information and its author's characteristic information in the root node may be weakened or forgotten in circulation. To address this, we introduce another root-aware attention module to learn the interactive correlations between the root node and each responsive node:
\begin{align}
	\widehat{\mathbf{V}}_i=\operatorname{Sigmoid}([\widetilde{\mathbf{V}}_0,\widetilde{\mathbf{V}}_i]\bm{W}^{a} + \mathbf{b}^{a}) * \widetilde{\mathbf{V}}_i
\end{align}
where $\bm{W}^{a} \in \mathbb{R}^{2d \times d}$ are projection weights, $\mathbf{b}^{a}\in \mathbb{R}^d$ is the bias term, $\widetilde{\mathbf{V}}_0$ and $\widetilde{\mathbf{V}}_i$ are the self-attention representations of the root node and the $i$-th responsive node, respectively, and $\widehat{\mathbf{V}}_i$ is the attention output attending to both the two nodes.

Finally, we fuse the outputs of the two attention modules:
\begin{align}
	\mathbf{U}_0&=\widetilde{\mathbf{V}}_0 + (\mu * V_0) \label{hyper1}
	\\\mathbf{U}_i&=\widetilde{\mathbf{V}}_i + (\mu * \widehat{\mathbf{V}}_i) \label{hyper2}
\end{align}
where $\mathbf{U}_0$ and $\mathbf{U}_i$ denote the fused representations of the root node and the $i$-th responsive node, respectively. In addition, $\mu$ is a hyper-parameter used to determine the extent of how much interactive correlation between the root node and responsive node should be considered. Subsequently, substituting $\mathbf{V}_i=\mathbf{E}_i$ or $\mathbf{V}_i=\mathbf{F}_i$ to the above Equations \ref{con:inatt}-\ref{hyper2} separately, we can obtain the representations of all post nodes and author nodes in the bipartite trees which are denoted as $\{\mathbf{S}_0, \mathbf{S}_1,...,\mathbf{S}_n\}$ and $\{\mathbf{O}_0, \mathbf{O}_1,...,\mathbf{O}_n\}$, respectively.

\subsection{Structural-level Representation Module}\label{sec:SRM}
All node-level representations learned above are fed into the \textit{Structural-level Representation module} to capture a comprehensive representation of both the post tree and author tree. The module consists of two sub-modules: \emph{Tree Variant RNN} (TRvNN) and \emph{Tree-aware Attention Learning} (TAL). TRvNN is a structure-aware GRU layer that simulates the hidden transitions according to the connections in the tree, while TAL aggregates the output hidden states from TRvNN to generate the representation vectors via a tree-aware attention mechanism.

\textbf{(1) Tree Variant RNN (TRvNN).} 
We incorporate the propagation tree topology to RNN to form TRvNN, a tree-variant RNN inspired by \cite{DBLP:conf/acl/WongGM18}. To guarantee computation efficiency in modeling the long propagation tree, we choose GRU cells with one node corresponding to one GRU cell.

Specifically, $\{\mathbf{S}_0, \mathbf{S}_1,...,\mathbf{S}_n\}$ and $\{\mathbf{O}_0, \mathbf{O}_1,...,\mathbf{O}_n\}$ are served as inputs to TRvNN. We assume that all information from a non-leaf node in the post tree or author tree can be passed to all of its descendant nodes. Therefore, the hidden state of the GRU cell is calculated by combining the representations of the current node and the hidden state of its direct parent node. Formally, given all node representations of the post tree or author tree $\{\mathbf{U}_0, \mathbf{U}_1,...,\mathbf{U}_n\}$, i.e., $\mathbf{U}_i=\mathbf{S}_i$ or $\mathbf{U}_i=\mathbf{O}_i$, the GRU cell in our TRvNN is calculated as follows:
\begin{align}
	\notag
	\widetilde{\mathbf{U}}_i&=\mathbf{U}_i\bm{W}^{U} + \mathbf{b}^{U}
	\notag
	\\
	\mathbf{r}_{i}&=\sigma_r([\widetilde{\mathbf{U}}_i,\mathbf{h}_{P(i)}]\bm{W}^r)
	\notag
	\\
	\mathbf{z}_{i}&=\sigma_z([\widetilde{\mathbf{U}}_i,\mathbf{h}_{P(i)}]\bm{W}^z) \label{con:RvNN}
	\\
	\widetilde{\mathbf{h}}_i&=\tanh([\widetilde{\mathbf{U}}_i,\mathbf{h}_{P(i)}*\mathbf{r}_i]\bm{W}^H)
	\notag
	\\\bm{h}_i&=(1-\mathbf{z}_i) * \mathbf{h}_{P(i)} + \mathbf{z}_i * \widetilde{\mathbf{h}}_i
	\notag
\end{align}	
where $P(i)$ is the parent of the $i$-th node in either the post tree or author tree, $\sigma_r,\sigma_z$ are the activation functions (precisely $\operatorname{Sigmoid}$ used in the experiments), $\bm{W}^{U}\in \mathbb{R}^{d\times d}$ and $b^{X}\in \mathbb{R}^{d}$ denote the transformation weights and the bias term, respectively, and $\mathbf{W}^r,\mathbf{W}^z,\mathbf{W}^H\in\mathbb{R}^{2d\times d}$ are the projection weights for calculating the reset gate $\bm{r}_i$, the update gate $\bm{z}_i$ and the candidate activation $\widetilde{\mathbf{h}}_i$, respectively. Accordingly, we can obtain all output hidden state $\{\bm{h}_i\}_{i=0}^n$ from TRvNN, resulting in $\{\bm{h}^p_i\}_{i=0}^n$ and $\{\bm{h}^a_i\}_{i=0}^n$ corresponding to all post nodes and author nodes, respectively.


\textbf{(1) Tree-aware Attention Learning (TAL).} To aggregate the output hidden state from TRvNN and capture structural information from the post/author tree, we propose a tree-aware attention learning module. This module leverages attention mechanisms and adopts $\{\bm{h}^p_i\}_{i=0}^n$ and $\{\bm{h}^a_i\}_{i=0}^n$ as inputs to obtain the final structural representations of the post tree and author tree. For simplicity, we introduce the module using $\{\bm{h}_i\}_{i=0}^n$, as shown in \myfigure{fig:myMod}(b).

First, based on the topological structure, we divide the obtained hidden state representations into three categories: the hidden states of the root node, leaf nodes, and other nodes, which are denoted by dotted boxes of diffident colors as shown in \myfigure{fig:myMod}(b). The treatment aims to show that different kinds of nodes in the post tree or author tree contribute differently to spreading rumors. That is, we distribute different weights to different kinds of nodes to learn tree-aware representations of the post/author tree:

(a) \emph{The leaf nodes} reflect the information aggregation process from the root node to the current leaf nodes along with the tree structure. Intuitively, the emotional intensity and propagation flow in the path is all gathered in the leaf nodes.

(b) \emph{The other nodes} act on the paths from the root node to the leaf nodes. Their node representations in the current path may influence the other paths. For example, the negative emotion in the current node may affect the expression of the nodes on other paths.

(c) \emph{The root node} contains the most important information in the propagation. However, during recursive learning, the information in the root node may be weakened or forgotten as the path length increases.

Following the above node division, all leaf nodes are fed into a max-pooling layer to obtain the maximum value of each vector element of all leaf nodes, thereby representing the representation of all leaf nodes. Intuitively, this treatment can account for and capture the most attractive and indicative representation of all propagation paths:
\begin{equation}
	\mathbf{h}_L = \operatorname{MaxPooling}(\mathbf{h}_{l1}, \mathbf{h}_{l2}, \mathbf{h}_{l3}\cdots)
\end{equation}
where $\mathbf{h}_{l1},\mathbf{h}_{l2},\mathbf{h}_{l3},\cdots$ denote the hidden states of the leaf nodes, the operator $\operatorname{MaxPooling}$ returns element-wise maximum values on $\{\mathbf{h}_{l1},\mathbf{h}_{l2},\mathbf{h}_{l3},\cdots\}$ over $\mathbb{R}^d$, and the resultant $\mathbf{h}_L\in\mathbb{R}^{d}$ denotes the maximum representation on all the leaf nodes. Accordingly, we calculate the correlation between this maximum representation and the root node representation and obtain an enhanced representation for the roof node $\mathbf{h}_E$:
\begin{equation}
	\mathbf{h}_E = \sigma_E([\mathbf{h}_0,\mathbf{h}_L]\bm{W}^E + \mathbf{b}^E) \label{enhance} 
\end{equation}
where $\mathbf{h}_0$ denote the hidden state representation of the root node, $\sigma_E$ is the activation function ($\operatorname{tanh}$ used in the experiments), $\bm{W}^E\in \mathbb{R}^{2d \times d}$ and $\mathbf{b}^E\in \mathbb{R}^{d}$ are the weight parameters and bias term, respectively. On the one hand, root node information can be fused with indicative maximum leaf node information to perform a strong combination via \myref{enhance}. On the other hand, it can compensate for the loss of root node information due to a long propagation.

Subsequently, the representation $\mathbf{h}_E$, along with other node hidden states $\{\bm{h}_i\}_{i=1}^n$, will be fed into a self-attention network. We expect to learn their correlations and capture the contribution of each node to the final tree structure representation. Let $\mathbf{h}_i$ be the $i$-th node hidden states, where $i \in [0, n]$ and $\mathbf{h}_0=\mathbf{h}_E$. We calculate the attentive weight $\alpha_i$ for each $\mathbf{h}_i$:
\begin{equation}
    \alpha_i=\frac{exp(\mathbf{h}_i\mathbf{w}_c^T)}{\sum_{j=0}^{n}{exp(\mathbf{h}_j\mathbf{w}_c^T})},
\end{equation}
where the parameter vector $\mathbf{w}_c\in\mathbb{R}^d$ can be treated as the learnable weight vector of a class token for rumor identification. Accordingly, we obtain the representation for the post tree or author tree via the attentive weighted summation:
\begin{equation}
    \label{equ:representation}
	\mathbf{h}_* = \sum_{i=0}^n \alpha_i\mathbf{h}_i
\end{equation}
Accordingly, we can obtain $\mathbf{h}_{post}$ and $\mathbf{h}_{author}$ according to Equation \ref{equ:representation} for the post tree and author tree, respectively.

\subsection{Prediction Layers and Loss Function}
To obtain comprehensive representations, we concatenate the representations of the post tree $\mathbf{h}_{post}$ and author tree $\mathbf{h}_{author}$, fusing the shared features of the two trees:
\begin{equation}
	\mathbf{h}_o = \mathbf{h}_{post} \oplus \mathbf{h}_{author}
\end{equation}
where $\mathbf{h}_o\in\mathbb{R}^{2d}$ denotes the output representation for prediction. Since we define the rumor detection task as a binary classification problem, we utilize a binary label $y\in [0,1]$ as the ground truth. To predict the label of the claim (the root node of the post tree), we feed $\mathbf{h}_o$ into a full-connection layer with the Softmax activation as follows:
\begin{equation}
	\hat{y} = \operatorname{Softmax}(\mathbf{h}_o\bm{W}^y + \bm{b}^y)
\end{equation}
where $\bm{W}^y\in\mathbb{R}^{2d\times 2}, \mathbf{b}^y\in\mathbb{R}^{2}$ are parameter weights and the bias term. Our model is trained to minimize the cross-entropy loss between each prediction $\hat{y}$ and the ground truth $y$ as below:
\begin{equation}
	\mathcal{L}(\Theta,\Omega) = -\sum\left[y\log \hat{y} + (1-y)\log(1-\hat{y})\right]
\end{equation}
where $\mathcal{L}$ denotes the cross-entropy loss, $\Theta=\{\{\mathbf{W}\},\{\mathbf{b}\}\}$ denotes the set of trainable weight parameters, and $\Omega=\{d,\mu\}$ is the set of hyperparameters.

\begin{table}[!t]
  \centering
  \caption{Statistics of the Datasets}
    \begin{tabular}{l|r|r}
    \toprule
    \midrule
    Statistic & PHEME & RumorEval \\
    \midrule
    Number of claim posts & 3,589  & 297 \\
    Number of authors & 57,961 & 4,261 \\
    Number of all posts & 61,782 & 4,841 \\
    Number of rumors & 1,803  & 137 \\
    Number of non-rumors & 1,786  & 160 \\
    Average number of reposts & 16.2 & 15.3\\
    Average tree depth & 4.6 & 4.1\\
    \midrule
    \bottomrule
    \end{tabular}%
  \label{datasets}%
\end{table}%

\section{Experiments} \label{Experiments}
\subsection{Experiment Details}
\textbf{Datasets.} We adopt two publicly available datasets, i.e., \emph{PHEME}\footnote{\url{https://figshare.com/articles/dataset/PHEME_dataset_of_rumours_and_non-rumours/4010619}}~\cite{2016Analysing} and \emph{RumorEval}\footnote{\url{https://alt.qcri.org/semeval2017/task8/index.php?id=data-and-tools}}~\cite{DBLP:conf/semeval/DerczynskiBLPHZ17}, for evaluation. Detailed statistics of these two datasets are provided in \myTable{datasets}. Both datasets are real-world data collected from the social platform Twitter, and the RumorEval dataset was developed for the SemEval-2017 Task 8 competition. The PHEME dataset contains five breaking news events, while RumorEval involves eight breaking news events. Both datasets contain post texts, author information, timestamps, and propagation information. Each claim is labeled with a binary class (i.e., True Rumor or False Rumor). In addition, during data preprocessing of the PHEME data, we removed noisy samples that have less than 3 retweets or do not have author information associated with the authors (IDs) of the claim post or retweet post. All comparative models are evaluated on the two datasets to provide an extensive and fair comparison.

\begin{table*}[!t]
  \centering
  \caption{Performance Comparison against Different Baselines on Two Datasets}
    \begin{tabular}{l|cccc|cccc}
    \toprule
    \midrule
    \multirow{2}[4]{*}{Method} & \multicolumn{4}{c|}{PHEME}    & \multicolumn{4}{c}{RumorEval} \\
\cmidrule{2-9}          & Acc. (\%) & Prec. (\%) & Rec. (\%) & F1 (\%) & Acc. (\%) & Prec. (\%) & Rec. (\%) & F1 (\%) \\
    \midrule
    DTR   & 58.30  & 58.56 & 58.07 & 58.25 & 56.10  & 53.66 & 61.97 & 57.52 \\
    CNN   & 59.23 & 56.14 & 64.64 & 60.09 & 61.90  & 54.54 & 66.67 & 59.88 \\
    TE    & 65.22 & 63.05 & 64.64 & 63.83 & 66.67 & 60.00    & 66.67 & 63.15 \\
    SVM   & 72.18 & 72.80  & 75.74 & 74.24 & 71.42 & 66.61 & 66.73 & 66.67 \\
    RvNN  & 81.75 & 81.24 & 82.46 & 81.85 & 81.48 & 78.49 & 85.32 & 81.77 \\
    DTCA  & 82.46 & 79.08 & 86.24 & 82.50  & 82.54 & 78.25 & 85.60  & 81.76 \\
    RvNN-GA & 85.19 & 84.93 & 86.21 & 85.56 & 86.03 & 85.08 & 86.51 & 85.79 \\
    BiGCN & 85.93 & 85.11 & 87.04 & 86.06 & 86.75 & \underline{87.79} & 85.73 & 86.75 \\
    ClaHi-GAT & 85.90  & -     & -     & 84.15 & -     & -     & -     & - \\
    RGNN & 86.17 & 85.98 & 86.29 & 86.57 & 87.03 & 86.87 & 87.51 & 86.79 \\
    EBGCN & 86.52 & \underline{86.85} & 86.38 & \underline{86.61} & 87.26 & 87.17 & 88.09 & \underline{87.63} \\
    FGCN & \underline{86.69} & 86.74 & \underline{87.07} & 86.56 & \underline{87.37} & 87.08 & \underline{88.26} & 87.49 \\
    BAET  & \textbf{87.88} & \textbf{87.41} & \textbf{89.24} & \textbf{88.31} & \textbf{88.65} & \textbf{88.23} & \textbf{89.58} & \textbf{88.87} \\
    \midrule
    \bottomrule
    \end{tabular}%
  \label{perfromance}%
\end{table*}

\textbf{Baselines}. We conduct a comprehensive comparison of our model with state-of-the-art rumor detection baselines.

\begin{itemize}
    \item \textit{DTR} \cite{DBLP:conf/www/ZhaoRM15}: A decision tree-based ranking model proposed to identify trending rumors through searching query phrases.
    
    \item \textit{CNN} \cite{DBLP:conf/semeval/ChenLK17}: A convolutional neural network with multiple filter sizes that is used to detect the stance of posts and determine the veracity of specific rumors.
    
    \item \textit{SVM} \cite{DBLP:conf/cikm/MaGWLW15}: A linear SVM classifier that is used to identify rumors by leveraging temporal sequences to model the chronological variation of social context features.
    
    \item \textit{TE} \cite{DBLP:conf/asunam/GuachoASP18}: A content-based model that leverages tensor decomposition to derive concise article embeddings, capture contextual information and create an article-by-article graph on limited labels.
    
    \item \textit{RvNN} \cite{DBLP:conf/acl/WongGM18}: A top-down (down-top) model based on tree-structured RvNN proposed to detect rumors through integrating structure and content. We choose the well-performed top-down model as our baseline.
    
    \item \textit{DTCA} \cite{DBLP:conf/acl/WuRZLN20}: A decision tree-based co-attention model developed to detect rumors by discovering evidence for explainable verification, in which the evidence scores combine both authors' and texts' information.
    
    \item \textit{RvNN-GA} \cite{DBLP:journals/tist/MaGJW20}: An enhanced rumor detection model proposed with designed discriminative attention mechanisms based on TD-RvNN/BU-RvNN. In this work, we select the top-down global attention model as a baseline.
    
    \item \textit{Bi-GCN} \cite{DBLP:conf/aaai/BianXXZHRH20}:  A bi-directional graph model based on two-layer GCNs to detect rumors through integrating structural features of both propagation depth (top-down) and dispersion (down-top).
    
    \item \textit{ClaHi-GAT} \cite{DBLP:conf/emnlp/LinMCYCC21}: A graph attention network model is proposed to represent posts and authors as an undirected graph and then augmented by a multi-layer attention mechanism to learn the rumor indicator characteristics.
    
    \item \textit{EBGCN}~\cite{DBLP:conf/acl/WeiHZYH20}: An edge-enhanced rumor detection model proposed with multiple graph conventional layers to iteratively aggregate features of structural neighbors and consider the uncertainty issue.
    
    \item \textit{RGNN}~\cite{9750388}: A rumor detection framework based on structure-aware retweeting graph neural networks to integrate content, users, and propagation patterns from a converted tractable binary tree to enhance performance.
    
    \item \textit{FGCN}~\cite{9837882}: A neuro-fuzzy method with fuzzy graph convolutional networks to capture uncertain interactions in the information cascade in a fuzzy perspective based on the tree or tree-like structure of source tweets.
\end{itemize}

These representative and state-of-the-art baselines are chosen carefully for extensive comparisons. 1) DTR, CNN, and SVM are representative baselines based on handcrafted features, and TE is a content-based detection model. Those baselines are compared to investigate the effectiveness of propagation information in the detection. 2) The other state-of-the-art baselines include RNN-, attention-, and GNN-based models that elaborately leverage propagation information to enhance detection. Those baselines are selected to investigate the benefits of our proposed bipartite adhoc trees, especially involving the propagation information from the author's perspective. In addition, multi-modality detection models are not considered in the experiments since those methods based on multi-modality information of text and image inputs do not belong to the family of propagation-based models.

\textbf{Experimental settings.} (1) \textit{Parameter settings}: In the experiments, all parameters are initialized with a normal distribution, and we perform a grid search of the learning rate over $\{1e^{-3}, 5e^{-3},1e^{-4}, 5e^{-4}, 1e^{-5}\}$ and L2 regularization rate over $\{0, 1e^{-1},1e^{-2},1e^{-3}, 1e^{-4}\}$ to guarantee the best performance of our proposed BAET. For hyperparameters $\Omega={d,\mu}$, we tune the dimension of the embedding and hidden units $d$ over $\{16, 32, 64, 128, 256\}$ and $\mu$ over $\{0.0,0.2,0.4,...,1.0\}$ to choose the values with the best performance for BAET. Finally, we set \romannumeral1) the embedding dimension $d=128$; ii) a balance weight $\mu=0.6$; iii) the ADAM optimizer~\cite{DBLP:journals/jmlr/DuchiHS11} with an initial learning rate of $0.005$; iv) a batch size of $16$; v) L2 regularization rate of $1e^{-3}$; vi) a dropout rate of $0.5$.
(2) \textit{Baseline settings}: Regarding the baselines, we implement RvNN, RvNN-GA, BiGCN, and EBGCN according to their published source codes, while other baselines are implemented by referring to the corresponding papers. To ensure fair comparisons, the best parameters for all comparative models are chosen by carefully tuning the author-recommended parameter ranges based on their recommended settings. Specifically, for the parameters in each baseline, we first initialize them with the values reported in the original paper and then carefully tune them on our datasets for best performance. 
(3) \textit{Metrics}: We adopt four metrics \textit{accuracy}, \textit{precision}, \textit{recall}, and \textit{F1-score} to evaluate all comparative rumor detection models. We conducted the experiments using \emph{five-fold cross-validation} and report the average evaluation results. (4) \textit{Platforms}: All experiments were implemented in PyTorch and were conducted on two NVIDIA 1080 12G GPUs.

\begin{figure}[!t]
	\centering
	\includegraphics[width=0.75\linewidth]{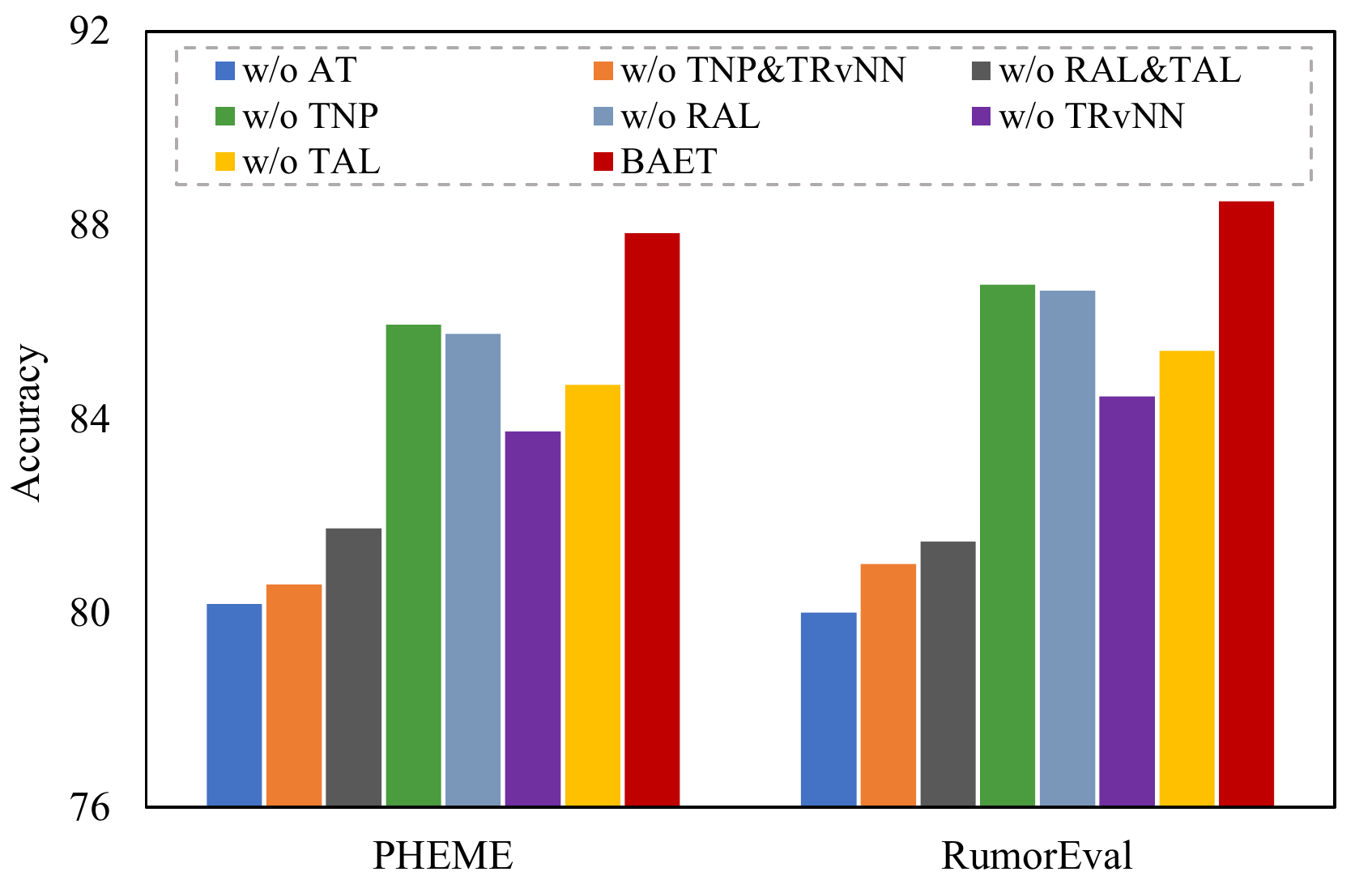}
	\caption{The post-based ablation analysis in terms of accuracy.}
	\label{fig:PostBased}
\end{figure} 

\subsection{Overall Experimental Performance}
In \myTable{perfromance}, we present the performance of different baseline models in comparison with our model based four evaluative metrics: \emph{accuracy}, \emph{precision}, \emph{recall} and \emph{F1-score}. The best results are highlighted in bold, and the best baseline results are underlined. From the table, we can make the following observations. 

(1) DTR, CNN, and SVM are rumor detection baselines based on handcrafted features. SVM can achieve superior results among methods based on handcrafted features. However, we attribute these results to the fact that SVM integrates temporal with textual information and, in doing so, simulates information transmission to a certain extent, which increases the relevance between texts.


(2) TE aims to mine rumor text features more thoroughly using social context. However, the table shows that the accuracy of the TE model is significantly lower (nearly 7\%) than that of the SVM model, indicating that multi-type feature fusion and information transmission topology may still be considered to achieve better performance.

(4) The proposed RvNN architecture was shown to provide a more convincing propagation tree structure for rumor detection. In this sense, RvNNs can better combine rumor propagation features and improve classification performance. On the other hand, the DTCA architecture achieved good performance by relying on a decision tree mechanism to detect rumors by establishing evidence scores while ignoring propagation information. RvNN-GA, with its unique attention mechanism, also achieves significant performance improvements compared with RvNN. However, these two models (RvNN, RvNN-GA) model the propagation topology solely on text contents, ignoring the characteristics of author groups involved in rumor spreading. Meanwhile, the exploration of interactive features between texts is insufficient.


(5) The proposed propagation tree structure broadens the scope of rumor detection research, and researchers focused on modeling the spread of the topological structure in more ways. BiGCN and EBGCN model the depth and breadth of propagation using graph convolution networks, which has yielded desirable results. Meanwhile, the EBGCN model proposes to investigate the uncertainty of propagation relations, which improves the model's interpretability.

(6) The baselines RGNN and FGCN also achieve desirable performance (the best two baselines). This is reasonable since RGNN and FGCN introduce GNN to capture the post text, author information, and propagation information. However, unlike RGNN and FGCN, BAET introducing bipartite adhoc event trees in terms of both posts and authors achieves better performance. The results demonstrate the effectiveness of introducing the author tree and modeling more comprehensive authors' information for rumor detection.

(7) On these datasets, we can see that our proposed BAET outperforms other benchmarks, proving that author information is effective for judging the veracity of the corresponding posts
"Birds of a feather propagate together" and our model structure is valid. Simultaneously, it is demonstrated that modeling propagation structure from the perspectives of texts and authors can mine the features gathered through propagation relations, and novel interactive learning modules are introduced to generate a more comprehensive representation of the claim to be detected.

\begin{figure}[t!]
	\centering
	\includegraphics[width=0.75\linewidth]{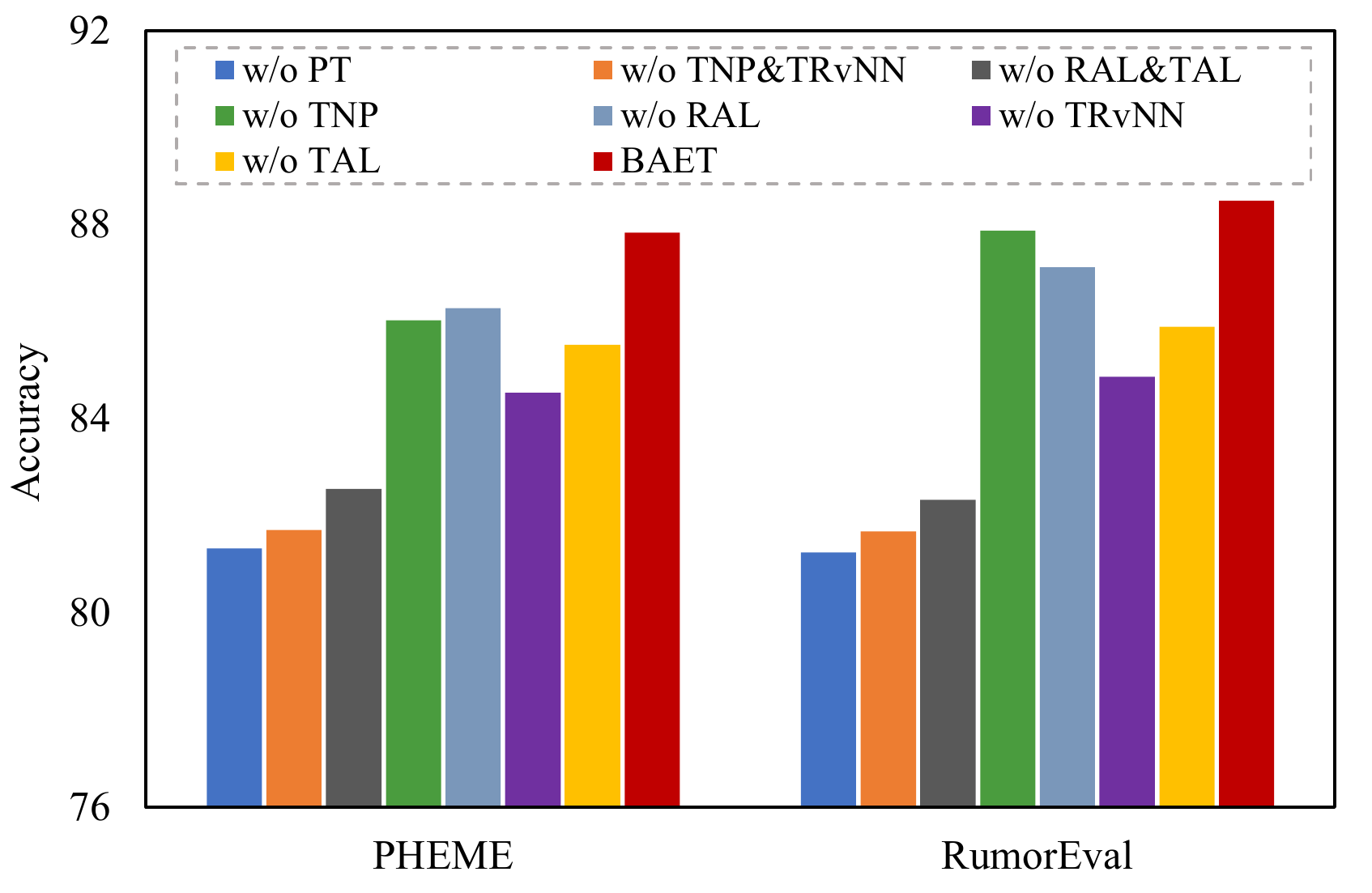}
	\caption{The author-based ablation analysis in terms of accuracy.} 
	\label{fig:AuthorBased}
\end{figure}

\subsection{Ablation Study}
To verify the contribution of each component of BAET, we perform ablation studies on PHEME and RumorEval, respectively, by discarding certain important components of BAET. Note that our BAET consists of two parallel components for the post tree and the author tree, respectively. Consequently, we perform the ablation study on the post-based and author-based components separately, denoted as post-based ablation and author-based ablation, respectively.

Both post-based and author-based components consist of two modules for node-level representation (including two successive sub-modules, TNP and RAL) and structural-level representation (including two successive sub-modules, TRvNN and TAL), which are necessary to learn to represent the bipartite adhoc event trees for detecting rumors. To ensure effective ablation studies, all variants of BAET keep the two-level structure, that is, at least one sub-module at either node-level or structure-level. Specifically, we compare BAET with its seven variants in the post-based and author-based ablation studies, respectively. We illustrate the variants in the post-based (author-based) ablation study as follows:

\begin{itemize}
    \item \textbf{w/o AT(PT)}: We discard the author (post) tree from BAET, namely using $\mathbf{h}_{post}$ ($\mathbf{h}_{author}$) only for detection.
    \item \textbf{w/o TNP\&TRvNN}: We remove TNP and TRVNN from the node-level representation and structural-level representation modules of BAET, respectively. Accordingly, only the two attention sub-modules, i.e., RAL and TAL, are maintained to learn node/structural representations.
    \item \textbf{w/o RAL\&TAL}: We remove the two attention sub-modules, i.e., RAL and TAL, from the node-level representation and structural-level representation modules of BAET, respectively, and feed the node embeddings from TNP to the tree-structural TRvNN to learn to detect.
    \item \textbf{w/o TNP}: We skip the sub-module TNP from the node-level representation module of BAET and feed the raw word embedding embeddings (author feature embeddings) into RAL to the node representations. 
    \item \textbf{w/o RAL}: We skip the sub-module RAL from the node-level representation module of BAET and feed the node embeddings from TNP directly into the structural-level representation module to detect.
    \item \textbf{w/o TRvNN}: We remove TRvNN from the structural-level representation module of BAET and feed the node representations into TAL to learn structural representations for detection.
    \item \textbf{w/o TAL}: We remove TAL from the structural-level representation module of BAET and output the last hidden state from TRvNN for detection.
\end{itemize}

We report the results of the post-based and author-based ablation studies in terms of accuracy in \myfigure{fig:PostBased} and \myfigure{fig:AuthorBased}, respectively. From the two tables, we observe that the proposed BAET outperforms all other variants in both datasets, indicating the contributions of each module to rumor detection and verifying the rationality and effectiveness of our model design in BAET. Specifically, we have the following observations.

(1) The accuracy of the variants \textbf{w/o AT} and \textbf{w/o PT} is inferior to that of BAET, demonstrating that modeling propagation structure using only posts or authors is insufficient. The results verify that the bipartite adhoc event trees comprehensively summarize the rumor propagation characteristics and that the circulation of post texts and author features contribute to modeling rumor propagation.

(2) The variants \textbf{w/o TNP\&TRvNN} and \textbf{w/o RAL\&TAL} fall behind BAET, indicating that TNP\&TRvNN and RAL\&TAL complement each other and contribute to modeling the propagation information in both the post tree and author tree. The attention modules (RAL\&TAL) help discover post (author) correlations in circulation and learn more informative representations, while TNP and TRvNN involve post/author features and reveal/embed the topological structure, respectively. In addition, the variant \textbf{w/o TNP\&TRvNN} performs worse than \textbf{w/o RAL\&TAL}, reflecting the significance of the propagation structure for rumor detection.

(3) The performance degradation of the remaining four variants compared with BAET, reflecting that the four sub-modules play different effects and complement each other for rumor detection. Specifically, TNP aims to model post text embeddings or author feature embeddings, RAL learns node representations attending to the root node, and TRvNN and TAL capture the topological structure and post (author) correlations, respectively. These designs effectively enhance the representations of rumor propagation from the node and structural levels.

(4) Finally, the performance of the variants in the \myfigure{fig:AuthorBased} is comparable with (slightly worse than) that of the variants in \myfigure{fig:PostBased}. The results indicate that the author characteristics in the author tree are beneficial for the task of rumor detection. In particular, the users' social attributes and writing habits are involved in learning author node representations. Empirically, these author features facilitate capturing authors' influence (such as reputation and reliability) on rumor propagation and significantly contribute to judging the veracity of rumors.

\subsection{Hyper-parameter Effect}

\begin{figure}[t!]
	\centering
	\includegraphics[width=0.75\linewidth]{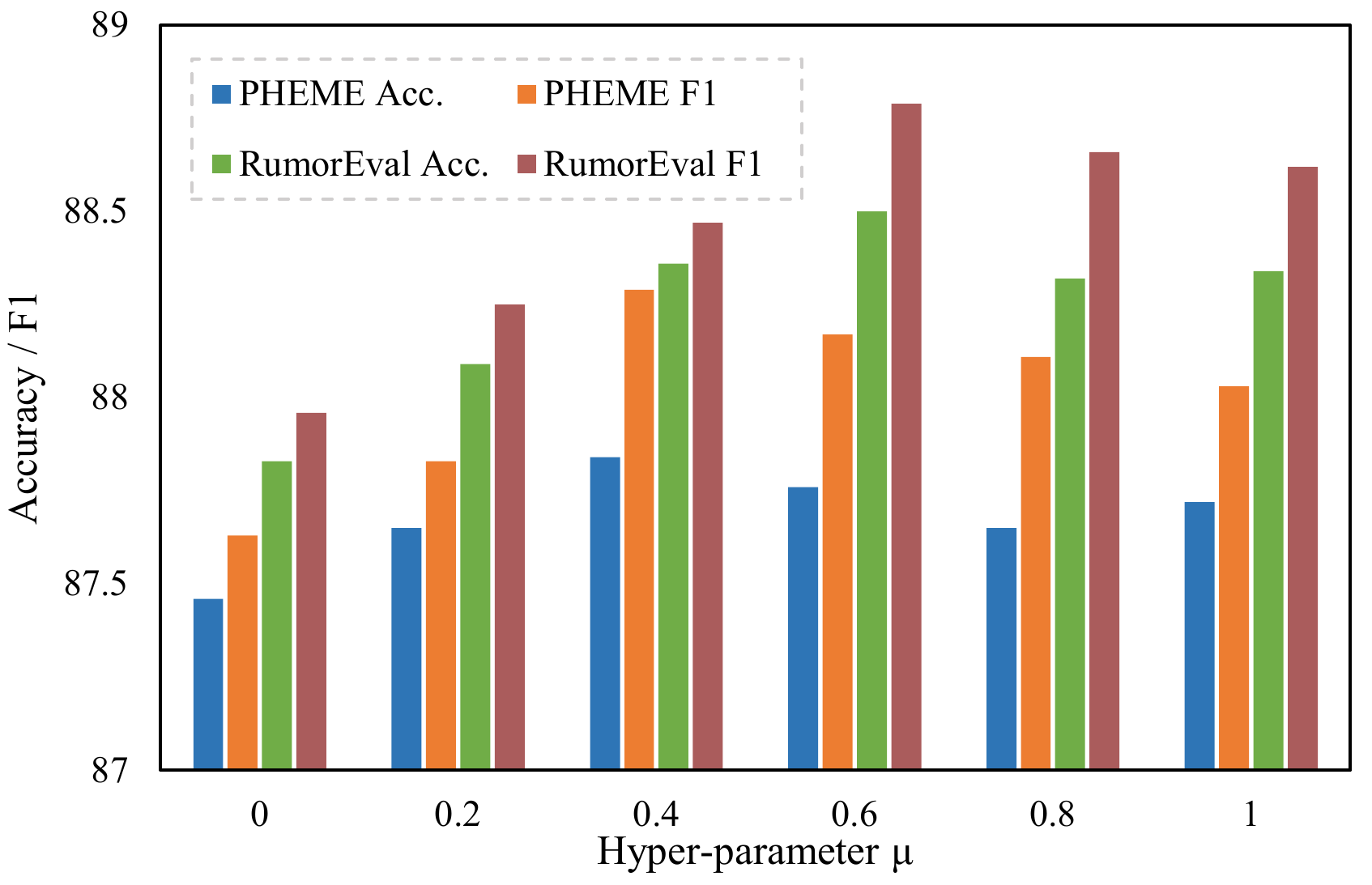}
	\caption{The accuracy of BAET under different hyper-parameter $\bm{\mu}$ on the two datasets.}\label{hyper}
\end{figure}

(1) First, we investigate the degree and difference to which information from the two interactive modes need to be considered in the \emph{Node-level Representation Module} in \myfigure{fig:myMod} (a) using hyper-parameters $\bm{\mu}$. To complete it, we use \myref{hyper1} and \myref{hyper2}. \myfigure{hyper} plots the Accuracy and F1 scores under different $\bm{\mu}$ values in \emph{Root-aware Attention Learning (RAL)}. From the figure, we can conclude that $\bm{\mu}$ of $0.4$ on PHEME and $0.6$ on RumorEval yield the best performance. According to the dataset introductions in \myTable{datasets}, we speculate that because the PHEME dataset contains a greater number of responsive posts, more propagation information can be captured, and thus the dependency degree of claim information from the claim is slightly lower than that of RumorEval. This demonstrates that the effects of responsive information and claim information on rumor detection results are complementary, and our model also effectively handles the fusion of the two types of them. Furthermore, when ignoring the root-aware attention layer, i.e., $\bm{\mu} = 0.0$, the performance of BAET gets worse, indicating that ignoring the correlations between the claim node and its successors yields a poor performance of BAET. The results verify the effectiveness and necessity of our proposed root-aware attention layer. Moreover, as $\bm{\mu}$ increases, the performance initially improves, while this does not mean that a bigger value of $\bm{\mu}$ is always better. If $\bm{\mu}$ exceeds a certain value, the model performance exhibits a downward trend, indicating that the correlation information obtained by two correlation modes may conflict with one another.

\begin{figure}[t!]
	\centering
	\includegraphics[width=0.7\linewidth]{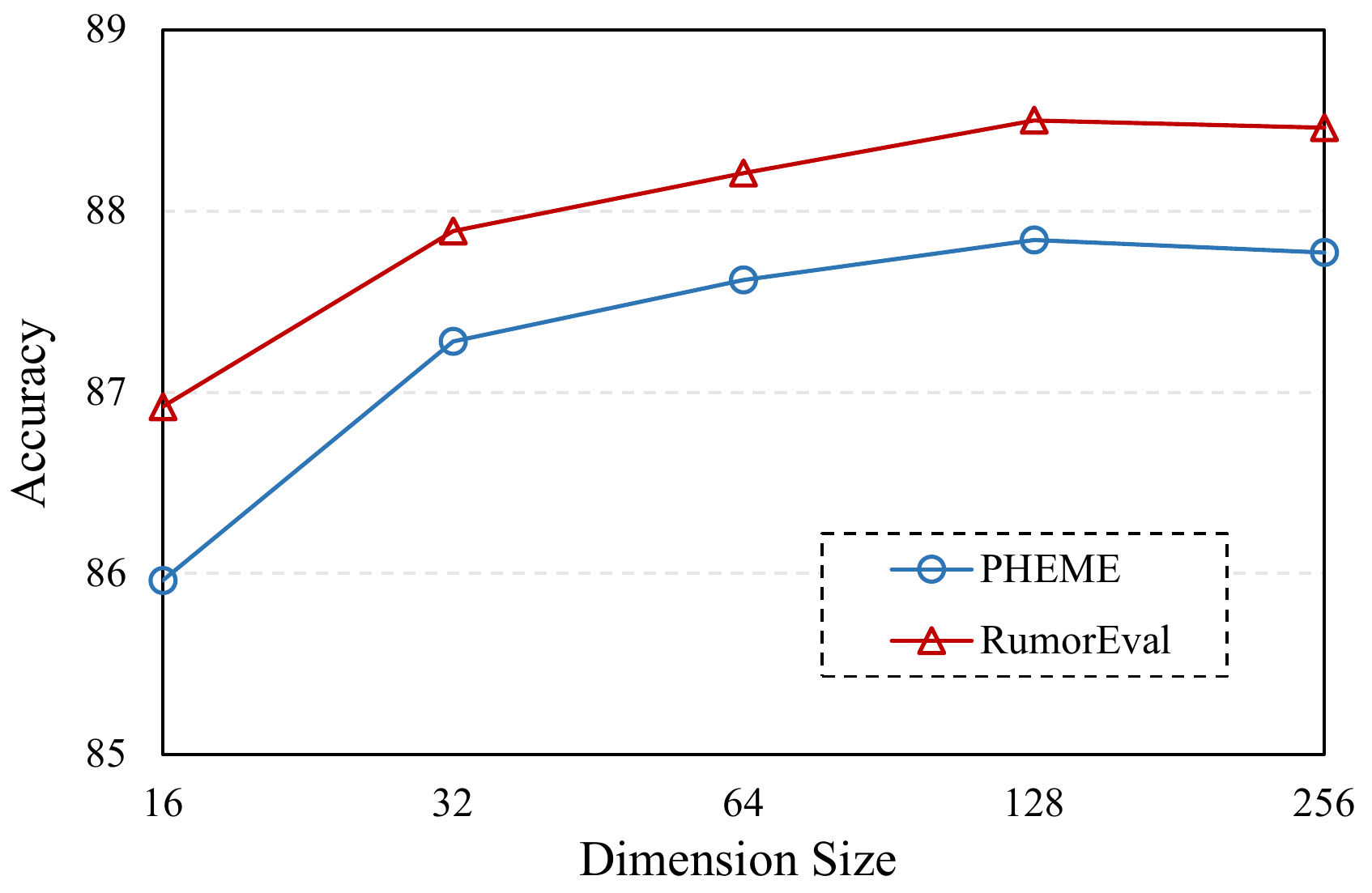}
	\caption{The accuracy of BAET under different dimension sizes $d$.} \label{dimension}
\end{figure}

(2) Moreover, \myfigure{dimension} shows the performance of our model under different dimension size $d$.  As shown, our model performance improves as dimension size $d$ increases. We hypothesize that increasing the embedding and hidden unit dimensions will result in more multidimensional and comprehensive representations of both post texts and author features and that these representations will provide richer information for model training. At the same time, the multi-hierarchical representation learning model that we developed is capable of capturing more interactive information between dimensions. Because our model achieves the best performance on both datasets when $d = 128$, we choose 128 as the dimension of embedding and hidden unit. When the dimension size, on the other hand, continues to grow, the model performance gradually improves or worsens. We believe that indefinitely increasing the representation dimension is ineffective. With the increasing dimensions, the model relies on more interactive dimension calculation information, which may increase the possibility of noise information and thus affect the model's detection effect.

\begin{figure}[t!]
	\centering
	\includegraphics[width=0.7\linewidth]{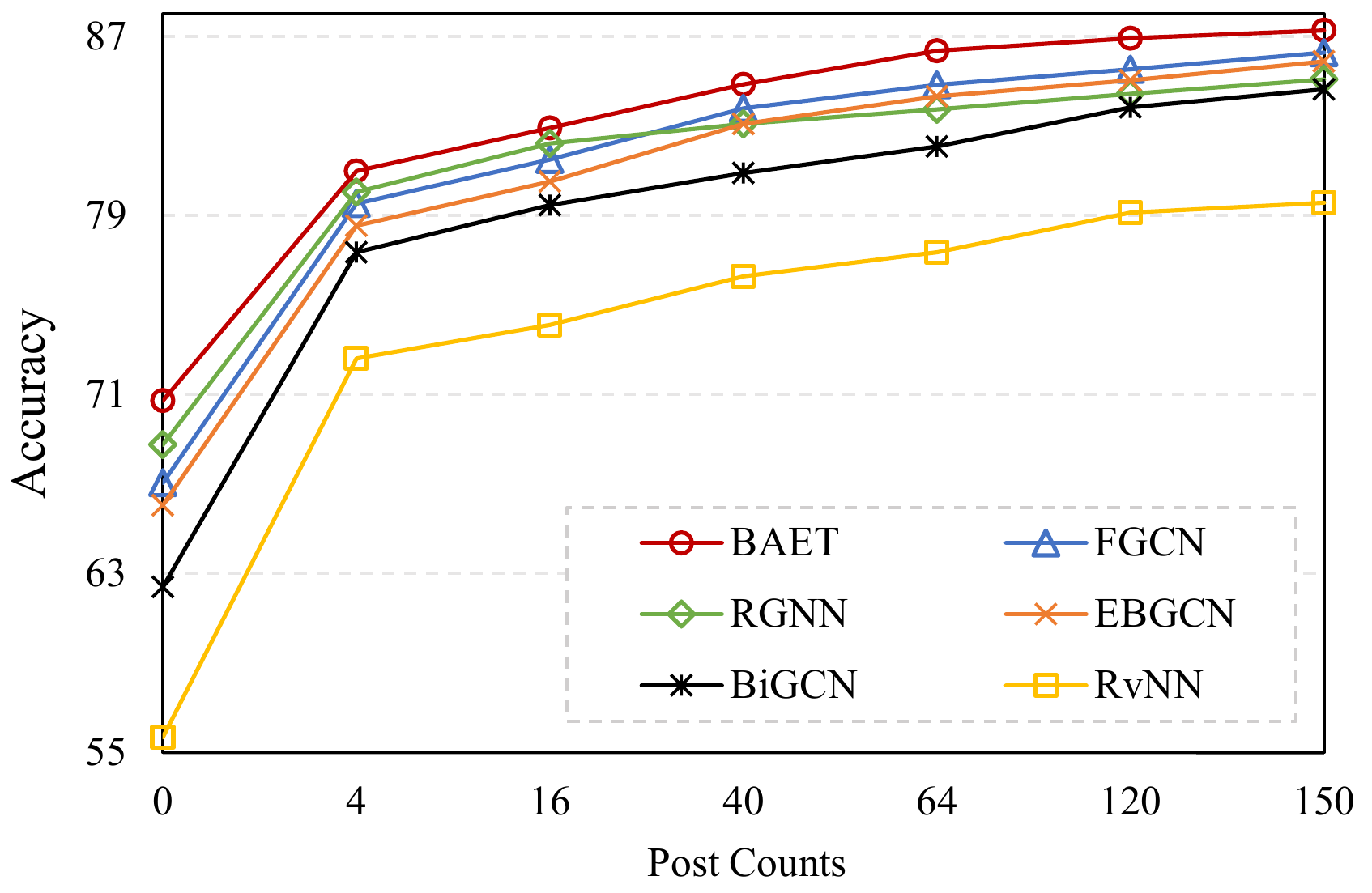}
	\caption{The accuracy under different responsive post counts on the PHEME.}\label{fig:post}
\end{figure}

\subsection{Post Count Effect} \label{PC-effect}
To further investigate the importance of the propagation process for rumor detection, we evaluate BAET with five propagation-based baselines (i.e., FGCN, RGNN, EBGCN, BiGCN, and RvNN) under the different number of retweets or forwarding (i.e., responsive post counts) to which a claim post. We report the accuracy of the comparative models on the PHEME dataset in \myfigure{fig:post}. As the number of posts increases from $0$ to $150$, our proposed BAET still outperforms all the baselines, indicating the superiority of BAET even for early rumor detection (i.e., few posts). When the number of posts is low, all models perform poorly, which is attributed to a lack of relevant information. The veracity of a claim cannot be determined immediately after it is sent out. During the propagation process, responding posts may express approval or disapproval of the claim, as well as provide some relevant evidence in the reply. The pieces of information are gathered through the propagation structure, which provides a solid foundation for identifying the rumor. 

As a result, as the number of posts grows, the performance of all models improves. Because multi-perspective modeling of rumor propagation and mining interactive information are taken into account, our model BAET outperforms all others in rumor detection. Second, RvNN proposed an investigation into the rumor detection propagation structure, but its detection capability is inferior to that of other models. Furthermore, while EBGCN and BiGCN have comparable performance, EBGCN has a better detection effect, and the BiGCN model has a more consistent performance increase. We hypothesize that the uncertainty of the propagation relation (edges in the tree) considered in the EBGCN model has an auxiliary effect on rumor detection, but because the complexity of the propagation relation increases with the number of posts, the EBGCN model's performance fluctuates slightly.

\begin{figure}[t!]
	\centering
	\includegraphics[width=\linewidth]{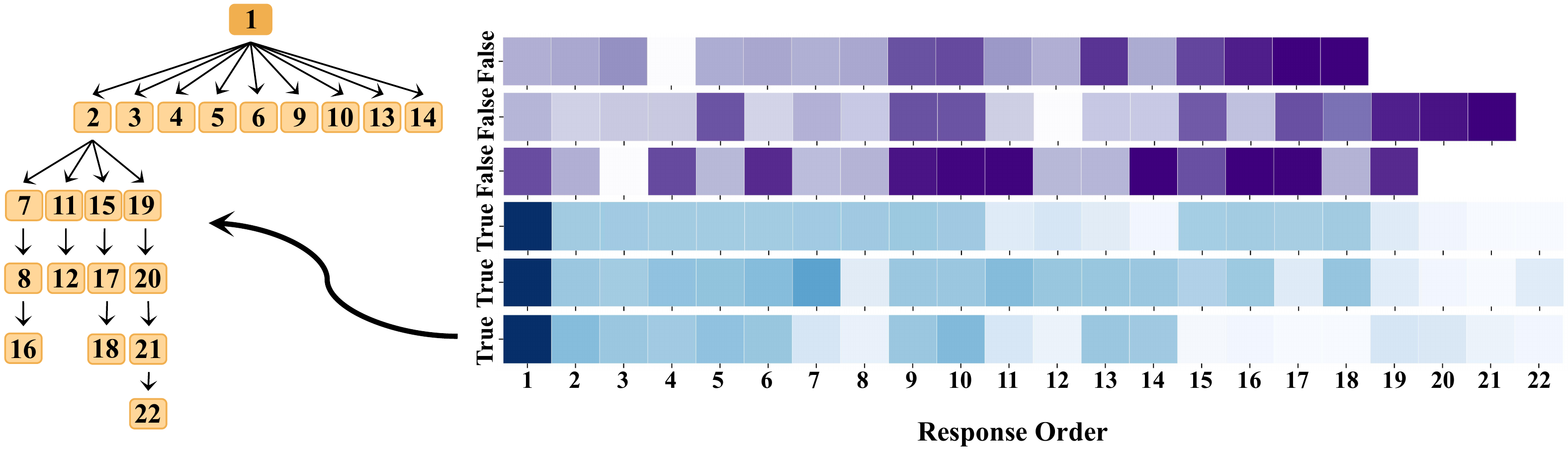}
	\caption{Visualization of attention weights for author propagation trees of three true and three false claims on \emph{PHEME} dataset. All nodes are placed from left to right in chronological response order. The darker the color is, the higher attention weight the corresponding node gets.}\label{visual}
\end{figure}

\subsection{Visualization Study}
To get a deep insight into the importance of authors who participate in rumors and non-rumors in the author propagation tree, we visualize the attention weights (\emph{Tree-aware Attention Learning} of claims randomly sampled from the \emph{PHEME} datasets. The result is shown in \myfigure{visual} with the author propagation tree structure of the first \emph{True Claim} in the left part. Three observations can be made from \myfigure{visual}: 

(1) Most authors have high weights, indicating that the author's propagation tree structure facilitates the modeling of claim veracity. 

(2) For true claims, authors closer to the root author yield a higher weight, i.e., these authors contain more features similar to the root author and "flock" around him, while authors form small-scale flocks during the propagation of false claims. This observation is confirmed by ~\cite{DBLP:journals/epjds/ZhaoZSLTTLWH20} that most people forward true rumors from a centralized source (claim), while rumors are transmitted by people forwarding to others. 

(3) Hence, we can infer that to judge a true claim, we could better check the traits of the early forwarding and closer authors. This is not absolute because the experiment in \myPara{PC-effect} shows that the number of responsive posts is also very important for rumor detection; thus, our model needs to be able to integrate multiple factors.

\section{Conclusion} 
\label{Conclusion}

In this study, we define the rumor propagation elements and use the propagation structure to represent the rumor claim as the special event tree structure. To fully exploit the above info, we propose a new rumor detection method (BAET) with hierarchical representation, which models the ad-hoc event tree into bipartite topological trees, namely (\emph{Author Tree} and \emph{Post Tree}) from two perspectives. And then "flocks" their respective features to obtain a better-integrated learning representation. Meanwhile, in the node-level representation module, we designed word embedding and feature encoder for both post and author separately; we also used root-aware attention learning to capture differentiated correlations between the tree nodes. Similarly, in the structural-level representation module, we adopted a tree-like RvNN network to learn structural dependencies, and tree-aware attention learning was used to mine attention correlations between the tree nodes. Eventually, the evaluation demonstrates that BAET is highly effective, outperforming state-of-the-art baselines.

\bibliographystyle{IEEEtran}
\bibliography{IEEEabrv,mybibfile}

\begin{IEEEbiography}[{\includegraphics[width=1in,height=1.25in,clip,keepaspectratio]{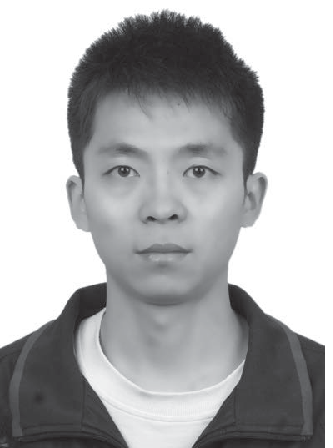}}]{Qi Zhang} received his Ph.D. degrees from Beijing Institute of Technology China under the dual Ph.D. program of Beijing Institute of Technology and University of Technology Sydney Australia. He is currently a research fellow at Tongji University. He has published high-quality papers in premier conferences and journals, including AAAI, IJCAI, SIGIR, TheWebConf, TKDE, TNNLS, and TOIS. His primary research interests focus on collaborative filtering, sequential recommendation, learning to hash, and MTS analysis. 
\end{IEEEbiography}

\begin{IEEEbiography}[{\includegraphics[width=1in,height=1.25in,clip,keepaspectratio]{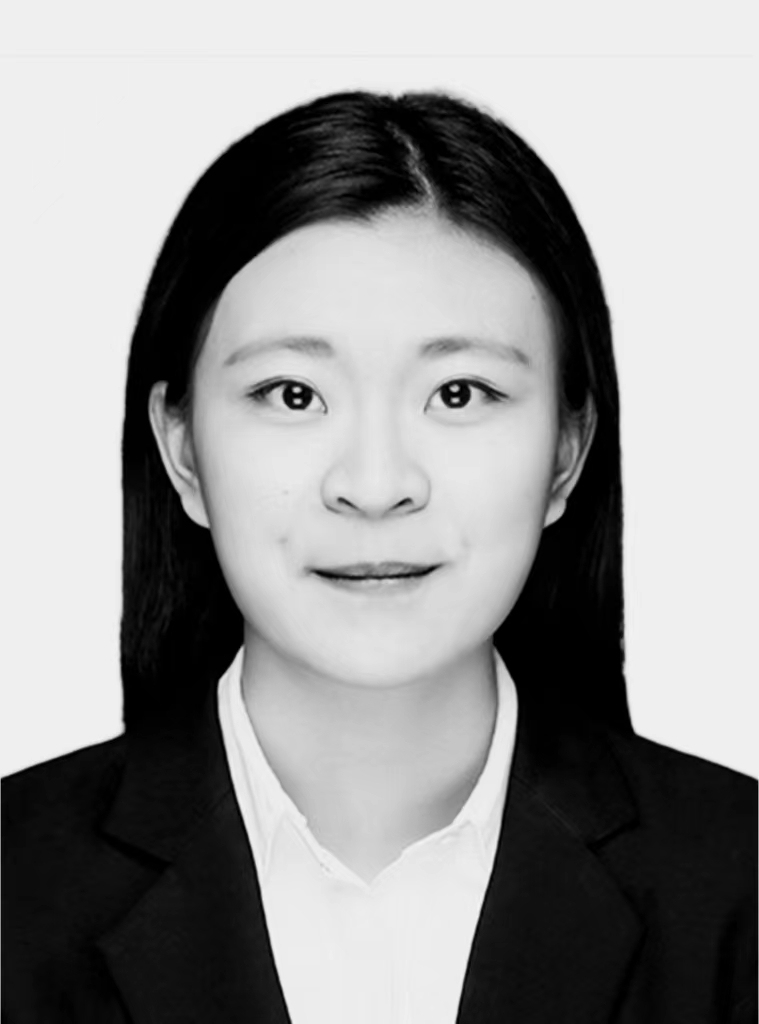}}]{Yayi Yang} is currently studying for her master's degree at Beijing Institute of Technology China in Computer Science. Her research interests focus on rumor detection, multi-modal fake news detection, and sentiment analysis.
\end{IEEEbiography}

\begin{IEEEbiography}[{\includegraphics[width=1in,height=1.25in,clip,keepaspectratio]{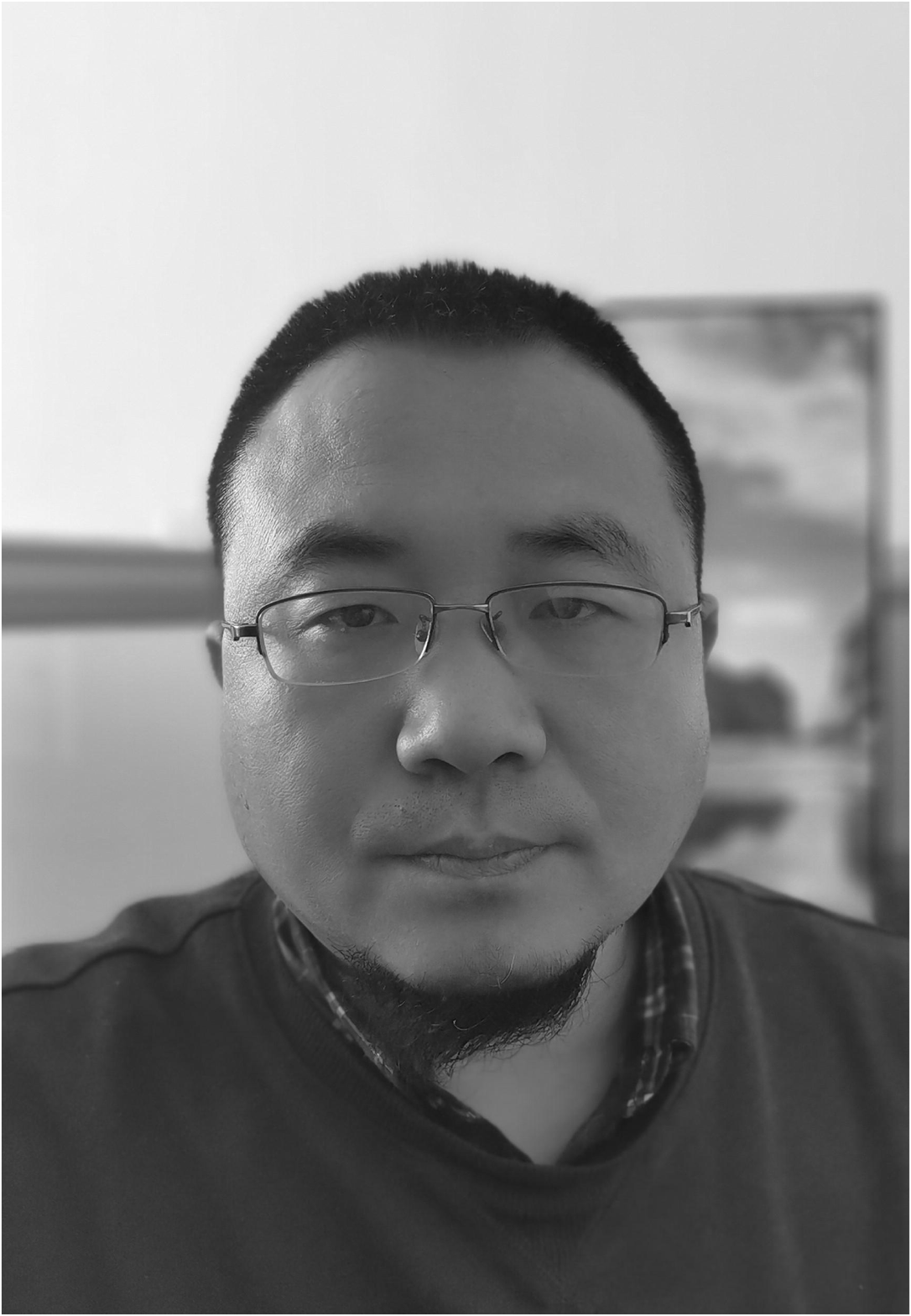}}]{Chongyang Shi} is currently an associate professor at the School of Computer Science, Beijing Institute of Technology. He received his Ph.D. degree from BIT in 2010 in computer science. Dr. Shi's research areas focus on information retrieval, knowledge engineering, personalized service, sentiment analysis, etc. He serves as an editorial board member for several international journals and has published more than 20 papers in international journals and conferences.
\end{IEEEbiography}

\begin{IEEEbiography}[{\includegraphics[width=1in,height=1.25in,clip,keepaspectratio]{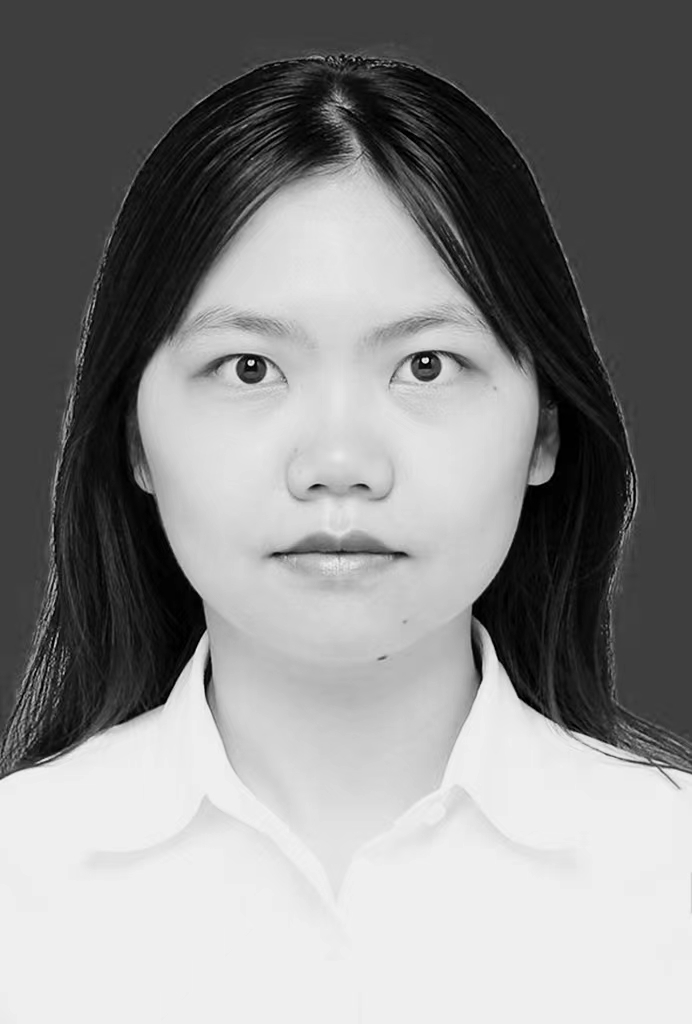}}]{An Lao} is studying her master's degree from Beijing Institute of Technology. Her primary research interests include rumor detention, sentiment analysis, and multi-modal fusion. She will pursue her Ph.D. at Beijing Institute of Technology.
\end{IEEEbiography}

\begin{IEEEbiography}[{\includegraphics[width=1in,height=1.25in,clip,keepaspectratio]{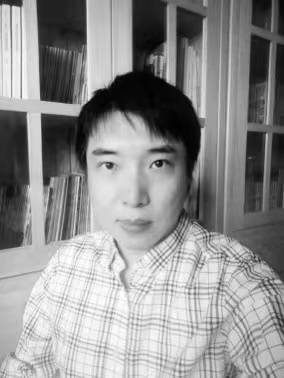}}]{Liang Hu} received dual Ph.D. degrees from Shanghai Jiao Tong University, China and University of Technology Sydney, Australia. He is currently a distinguished research fellow at Tongji University and chief AI scientist at DeepBlue Academy of Sciences.
His research interests include recommender systems, machine learning, data science, and general intelligence. 
\end{IEEEbiography}

\begin{IEEEbiography}[{\includegraphics[width=1in,height=1.25in,clip,keepaspectratio]{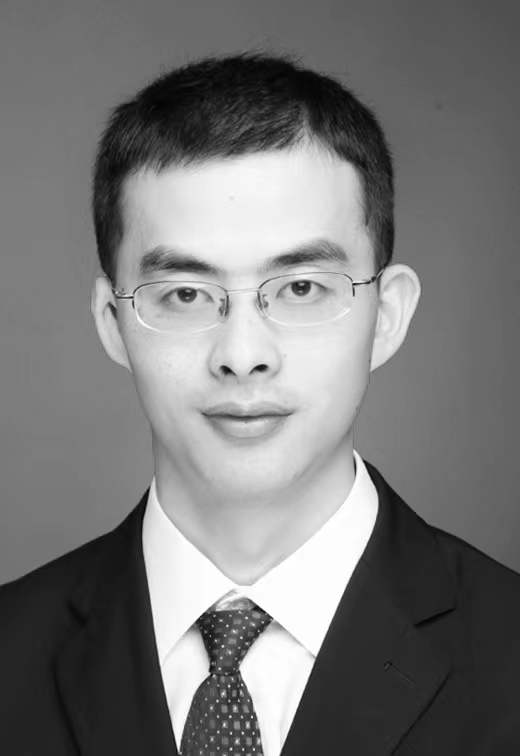}}]{Shoujin Wang} received a Ph.D. in Data Science from University of Technology Sydney (UTS) in 2019. He is currently a Lecturer in Data Science at UTS, Australia. 
His research interests include data mining, machine learning, recommender systems, and fake news mitigation. He has published high-quality papers in premier conferences and journals, including TheWebConf, AAAI, IJCAI, ECML-PKDD, and ACM CSUR. He is a recipient of some prestigious awards, including the 2022 DSAA Next-generation Data Scientist Award and the 2022 Club Melbourne Fellowship Award.  
\end{IEEEbiography}

\begin{IEEEbiography}[{\includegraphics[width=1in,height=1.25in,clip,keepaspectratio]{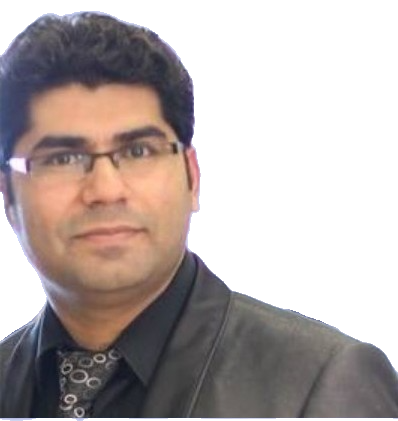}}]{Usman Naseem} is a Ph.D. student in the School of Computer Science, University of Sydney, Australia. He received his master’s degree in Analytics (research) from the School of Computer Science, University of Technology Sydney, Australia, in 2020. His primary research is in the intersection of machine learning and natural language processing for social media analytics and biomedical/health informatics.
\end{IEEEbiography}

\end{document}